\documentclass[aps,prb,twocolumn,amsmath,amssymb,superscriptaddress,floatfix]{revtex4-1}

\makeatletter
\def\l@subsubsection#1#2{}
\makeatother

\usepackage{amsfonts}
\usepackage{listings}
\usepackage{enumerate}
\usepackage{latexsym}
\usepackage{bm}
\usepackage{graphicx,float}
\usepackage{siunitx}
\usepackage{subfigure}
\usepackage{braket}
\usepackage{xcolor}
\usepackage{verbatim}
\usepackage{times,tikz}

\usepackage{ulem} 
\newcommand{\bs}[1]{\boldsymbol{#1}}

\usepackage{bbold}
\usepackage{hhline}

\begin{document}
     
\widowpenalty10000
\clubpenalty10000

\title{
Fractional corner charges in spin-orbit coupled crystals}

\author{Frank Schindler}
\affiliation{Department of Physics, University of Zurich, Winterthurerstrasse 190, 8057 Zurich, Switzerland}
\affiliation{Kavli Institute for Theoretical Physics, University of California, Santa Barbara, CA 93106, USA}

\author{Marta Brzezi\'{n}ska}
\affiliation{Department of Theoretical Physics, Faculty of Fundamental Problems of Technology, Wroc\l{}aw University of Science and Technology, 50-370 Wroc\l{}aw, Poland}
\affiliation{Department of Physics, University of Zurich, Winterthurerstrasse 190, 8057 Zurich, Switzerland}

\author{Wladimir A. Benalcazar}
\affiliation{Department of Physics, The Pennsylvania State University, University Park, PA 16802, USA}

\author{Mikel Iraola}
\affiliation{Donostia International Physics Center, 20018 Donostia-San Sebastian, Spain}
\affiliation{Department of Condensed Matter Physics, University of the Basque Country UPV/EHU, Apartado 644, 48080 Bilbao, Spain}

\author{Adrien Bouhon}
\affiliation{Department of Physics and Astronomy, Uppsala University, Box 516, SE-751 21 Uppsala, Sweden}
\affiliation{NORDITA, Roslagstullsbacken 23, 106 91 Stockholm, Sweden}

\author{Stepan S. Tsirkin}
\affiliation{Department of Physics, University of Zurich, Winterthurerstrasse 190, 8057 Zurich, Switzerland}

\author{Maia G. Vergniory}
\affiliation{Donostia International Physics Center, 20018 Donostia-San Sebastian, Spain}
\affiliation{IKERBASQUE, Basque Foundation for Science, Maria Diaz de Haro 3, 48013 Bilbao, Spain}
            
\author{Titus Neupert}
\affiliation{Department of Physics, University of Zurich, Winterthurerstrasse 190, 8057 Zurich, Switzerland}

\date{\today}

\begin{abstract}
We study two-dimensional spinful insulating phases of matter that are protected by time-reversal and crystalline symmetries. To characterize these phases we employ the concept of corner charge fractionalization: corners can carry charges that are fractions of even multiples of the electric charge. The charges are quantized and topologically stable as long as all symmetries are preserved. We classify the different corner charge configurations for all point groups, and match them with the corresponding bulk topology. For this we employ symmetry indicators and (nested) Wilson loop invariants. We provide formulas that allow for a convenient calculation of the corner charge from Bloch wave functions and illustrate our results using the example of arsenic and antimony monolayers. Depending on the degree of structural buckling, these materials can exhibit two distinct obstructed atomic limits. We present density functional theory calculations for open flakes to support our findings.
\end{abstract}

\date{\today}

\maketitle

\section{Introduction}

The classification of insulating phases of matter by topological invariants has been refined in important ways recently. 
A detailed understanding of symmetries is paramount for the classification of topological phases.
An example foundational to the field are the $\mathbb{Z}_2$ topological insulators in two dimensions (2D) and three dimensions (3D)~\cite{Kane05a,Kane07,z2spinpumpfukane,Fu07}, whose topology is protected by time-reversal symmetry (TRS) and manifests in edge and surface states, respectively, which are immune to Anderson localization. Two characteristics of these phases can equivalently be used to define them as \emph{topological}: (i) the bulk cannot be adiabatically deformed into the trivial phase (an atomic limit or the vacuum) while retaining the protecting symmetry and (ii) boundary modes protected by the symmetry appear. 
  
Recently, spatial and in particular space group symmetries have been used to define topological properties~\cite{Berg08,PollmannSPTprecursor10,Fu11,LevinGuSPT12,GroupCohomologyPaper13,Ando15,Hermele17PointGroups}, characterizing topological crystalline insulators. As this extends the number of known topological phases significantly, it also calls for a sharper definition of what is \emph{topological}, as the criteria (i) and (ii) do not coincide anymore when spatial symmetries are required for topological protection. For one, the topological bulk-boundary correspondence is extended to include higher-order topological insulators (HOTIs) which exhibit hinge or corner modes in 3D and corner modes in 2D~\cite{BABHughesBenalcazar17,Brouwer17,SchindlerHOTI,Benalcazar17,Khalaf18,Geier18,BenMoTe2,wang2019two}. Secondly, the atomic limit as a trivial reference point is not unique~\cite{Bradlyn17}. With spatial symmetries, several atomic limits exist that cannot be adiabatically deformed into one another. The physical reference point for an actual material is the atomic limit that corresponds to the physical location of ions. A situation where Wannier charge centers of the occupied bands of an insulator correspond to a different atomic limit is referred to as an obstructed atomic limit~\cite{Bradlyn17} (OAL).

This leaves three types of (spatial) symmetry-protected phases that can be distinguished according to the criterion of bulk phase transitions: (1) phases which cannot be adiabatically transformed to any atomic limit, which we refer to as strong topological, (2) phases which correspond to an OAL. Curiously, there is a third category of  
(3) phases with fragile topology~\cite{AshvinFragile,BarryFragile,Bouhon18Fragile,KoreanTBG,wieder2018axion,MaiaFragile1,MaiaFragile2,MaiaFragile3}, which are not adiabatically deformable to an atomic limit, but can be deformed to one upon the addition of bands that correspond to an atomic limit.
Many of the phases of type (2) support point-like boundary states in an open geometry, since the physical boundary of the system may ``cut through'' the Wannier charge centers in the OAL. At fixed bulk filling, these ``dangling'' Wannier charge centers become fractionally filled, which allows to define fractional charges, even in the absence of accompanying boundary states~\cite{miertOrtixFractionalCharge17,miertcorners,EzawaWannier19,benalcazar2018quantization,lee2019higher,sheng2019two}. 
The notion of a \emph{filling anomaly}\cite{benalcazar2018quantization} makes this precise: at a filling that corresponds to an insulating band gap with periodic boundary conditions, the system has to be metallic with open boundary conditions when the relevant symmetries are respected, because the boundary modes are then fractionally filled. This notion does not require a spectral symmetry, which is often used to pin boundary modes in the middle of a band gap. 
A paradigmatic example are one-dimensional (1D) lattices with reflection (inversion) symmetry, as represented by the Su-Schrieffer-Heeger model, for instance. There exist two atomic limits, with Wannier centers at inequivalent high-symmetry positions in the unit cell. One of them, with Wannier center at the unit cell boundary (potentially realized in polyacetylene) leads to half charges at the end of an open chain~\cite{Miert16,miertOrtixFractionalCharge17,rhim17}. However, it is only with spectral symmetries imposed that systems with end modes are strong topological phases. 

Obstructed atomic limits in 2D are the natural extensions of the above-mentioned topological phases in 1D. Two-dimensional lattices, however, can exhibit richer classifications: even in the absence of bulk polarization, point-like corner charges can get generated~\cite{BABHughesBenalcazar17,Benalcazar17,Song17}. From the exclusive perspective of charge fractionalization, these cases -which are protected only by spatial symmetries- broaden our understanding of second-order topological insulators~\cite{BABHughesBenalcazar17,Brouwer17,SchindlerHOTI,Benalcazar17, Song17,Khalaf18,Geier18,BenMoTe2}, which have the additional requirement of particle-hole or chiral symmetries. The relation between 2D Wannier centers and corner charge was developed in Refs.~\onlinecite{Song17,miertcorners,benalcazar2018quantization}. In particular, Ref.~\onlinecite{benalcazar2018quantization} uses the algebraic structure of the classifications of $C_n$-symmetric insulators in class AI (spinless, time-reversal symmetric insulators) to build topological indices for corner charge. It has also been recently found that fragile topological phases can also host corner charges~\cite{benalcazar2018quantization,wieder2018axion} and that 2D second-order topological insulators also exhibit fractional charges at the core of defects with curvature singularities in both spinless and spinful insulators~\cite{benalcazar2018quantization,liu2019,li2019fractional}.

In this work, we are concerned with 2D TRS spin-orbit coupled crystalline solids that admit a band structure description in terms of free fermions and fall in category (2) above, i.e., OALs. (This excludes in particular $\mathbb{Z}_2$ topological insulators protected by TRS and phases with a mirror Chern number, where the mirror plane is the plane of the 2D solid itself.) Our aim is to classify OALs with fractional corner charges/filling anomalies in all 2D layer groups and to provide topological invariants which allow to infer the presence or absence of such charges from the knowledge of the bulk band structure alone. Such invariants are either computed from the irreducible representations of the Bloch states -- in which case we speak of \emph{symmetry indicators} -- or expressed as integrals of a connection obtained from the Bloch states over (subsets of) the Brillouin zone, which will be referred to as Berry phase or Wilson loop type invariants. In particular, we make use of nested Wilson loop invariants~\cite{BABHughesBenalcazar17} which were particularly helpful in diagnosing OALs for which symmetry indicators fall short. With the invariants presented here, we can identify OALs in a computationally more efficient way than by explicitly computing maximally localized Wannier functions. Importantly, the corner charge/filling anomaly of such a 2D system cannot be removed by any symmetry-respecting boundary manipulation, including the ``gluing'' of additional degrees of freedom to the boundary.

In general, a set of occupied bands (without strong or fragile topology) can be decomposed into subsets of bands stemming from localized orbitals at different Wyckoff positions. The minimal subblocks that cannot be further decomposed are \emph{elementary band representations} (EBRs), which are a connected set of subbands induced from placing a certain orbital at a given Wyckoff position~\cite{Slager12,Slager17,Bradlyn17,Cano17,Cano17-2,BarryFragile,ZakEBR1,ZakEBR2}. Reference~\onlinecite{Bradlyn17} introduced EBRs as a means to discern bands that stem entirely from atomic limits from those with strong or fragile topology. Following the approach in Ref.~\onlinecite{benalcazar2018quantization}, here we use the additive structure of atomic limits that they provide to establish the correspondence between bulk invariants and corner charges sourced by OAL Wannier charge centers.

In addition to the general classification, we discuss buckled bilayers of Bi, As, and Sb as a family of 2D materials that can realize OALs and corner charges. Based on density functional theory (DFT) calculations, we provide a phase diagram as a function of buckling strength (which may be controlled with a suitable substrate), identifying strong topological insulator (TI) phases and 2D OALs. One of the OALs supports corner charges, while the other one does not.

Our paper is structured as follows. In Sec.~\ref{sec: charge quantization} we define the precise meaning of corner charge for our work as well as the role of the sample termination. In Sec.~\ref{sec: topo invariants}, we show how OALs can be identified by bulk invariants. Finally, in Sec.~\ref{sec:materials} the material candidates are discussed.

\section{Charge quantization in time-reversal symmetric spin-orbit coupled insulators}
\label{sec: charge quantization}
The 2D phases we are interested in support no 1D gapless boundaries. The bulk gap may be populated by boundary-localized midgap states, but those cannot be stabilized by TRS or crystalline symmetries. In this section, we will show that it is nevertheless often possible to diagnose a phase as an OAL in 2D via its corner charge fractionalization. We establish that this property remains invariant even when all midgap states are pushed out of the bulk gap and arbitrary symmetry-preserving boundary manipulations are allowed.~\cite{Benalcazar17,Ezawa18,miertcorners,KoreanTBG,benalcazar2018quantization} 

\begin{figure}[t]
\centering
\includegraphics[width=0.98\columnwidth]{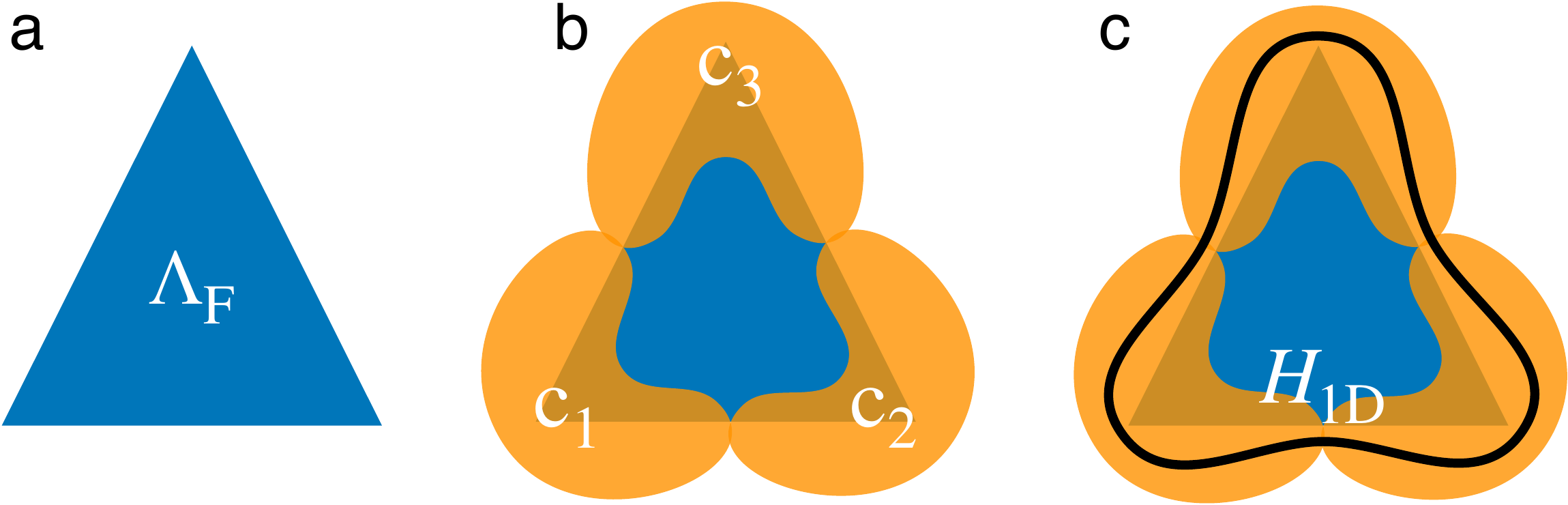}
\caption{Corner charge fractionalization due to $C_3$ rotational symmetry. \textbf{a}~The finite system $\Lambda_\mathrm{F}$ on which $H_\mathrm{F}$ is defined. \textbf{b}~The boundary regions $\mathrm{c}_1, \mathrm{c}_2, \mathrm{c}_3 \in \mathrm{C}$. Due to the $C_3$ symmetry in $\mathcal{G}_\mathrm{F}$, we have that $Q_{\mathrm{c}_1} = Q_{\mathrm{c}_2} = Q_{\mathrm{c}_3}$. Together with $Q_{\mathrm{c}_1} + Q_{\mathrm{c}_2} + Q_{\mathrm{c}_3} \in 2 \mathbb{Z}$, this implies a corner charge fractionalization in even multiples of $1/3$. \textbf{c}~A 1D edge addition, modeled by the Hamiltonian $H_{\mathrm{1D}}$. We prove in section~\ref{sec:bdrychargerobustness} that the corner charges $Q_{\mathrm{c}_i}$ are only changed by even integers.}
\label{fig:cornerchargefrac}
\end{figure}

\subsection{Quantization of the corner charge}
\label{sec:bdrychquant}
We consider 2D spinful insulating systems with TRS $T$ (class AII in the Altland-Zirnbauer classification, $T^2 = -1$) and the spatial symmetries corresponding to a symmetry group $\mathcal{G}$. We exclude first-order topological insulators so that the models we are studying are generically gapped even in a geometry with open boundary conditions. Additionally, we exclude insulators with bulk (TRS) polarization because those have edge-induced filling anomalies that scale with edge length and therefore result in metallic edges that preclude the existence of stable localized corner charges\cite{benalcazar2018quantization}.

We assume a tight-binding description of the system of interest. Denote by $\bs{a}_1$ and $\bs{a}_2$ the translation vectors corresponding to the decomposition of the $\mathcal{G}$-symmetric lattice $\Lambda$ into $n$-site unit cells $\mathrm{S} = \{\bs{r}_1, \dots, \bs{r}_n \}$, where $\bs{r}_i$ denotes the position of site $i$ in the unit cell as measured from the unit cell origin $\bs{r}_1 \equiv \bs{0}$. (Note that here and in the following, we only treat unit cells that are mapped to themselves under all available point-group symmetries, and do not cut through atomic sites. By these properties, a finite-size termination which does not cut through unit cells becomes possible.) We have
\begin{equation}
\Lambda = \bigcup_{x,y \in \mathbb{Z}} \bigcup_{\bs{r} \in \mathrm{S}} \left(x \bs{a}_1 + y \bs{a}_2 + \bs{r} \right).
\end{equation}
We are considering tight-binding Hamiltonians of the form
\begin{equation}
\label{eq:tbham}
H = \sum_{\bs{v}, \bs{w} \in \Lambda} \sum_{\mu,\nu} h_{\bs{v} \mu, \bs{w} \nu} c^\dagger_{\bs{v} \mu} c^{\vphantom{\dagger}}_{\bs{w} \nu},
\end{equation}
where $\mu,\nu$ run over orbital degrees of freedom defined at each lattice site and $c^\dagger_{\bs{v} \mu}$ creates an electron in orbital $\mu$ at lattice site $\bs{v}$. Hermiticity of $H$ as well as the symmetry requirements posed by $T$ and the symmetry group $\mathcal{G}$ imply relations among the Hamiltonian elements $h_{\bs{v} \mu, \bs{w} \nu}$ which we implicitly assume to be fulfilled here and in the following.

Given a unit cell decomposition of $\Lambda$ in terms of $S$, we define a \emph{trivial atomic limit} by a Hamiltonian that is adiabatically deformable into one for which the implication
\begin{equation}
\quad \bs{v} \in \bigcup_{\bs{r} \in \mathrm{S}} \left(x \bs{a}_1 + y \bs{a}_2 + \bs{r} \right) \not\ni \bs{w} \quad \Rightarrow \quad h_{\bs{v} \mu, \bs{w} \nu} = 0
\end{equation}
holds for all choices of $x$ and $y$, that is, there are no couplings between different unit cells.

To calculate corner charges we consider a finite system of $|\mathrm{F}|$ unit cells, via restricting $H$ to a subset $\Lambda_\mathrm{F} \in \Lambda$ (thereby obtaining $H_\mathrm{F}$), which is given by
\begin{equation}
\label{eq:finitelattice}
\Lambda_\mathrm{F} = \bigcup_{x,y \in \mathrm{F}} \bigcup_{\bs{r} \in \mathrm{S}} \left(x \bs{a}_1 + y \bs{a}_2 + \bs{r} \right).
\end{equation}
We choose $\Lambda_\mathrm{F}$ so as to retain all point group symmetries contained in $\mathcal{G}$, a subgroup we denote by $\mathcal{G}_\mathrm{F}$ (it does not contain translations or nonsymmorphic symmetries). Then we consider a subset $\mathrm{C} \subset \Lambda_\mathrm{F}$ comprised of a minimal (but larger than 1) number of disjoint boundary regions that form an orbit under $\mathcal{G}_\mathrm{F}$ and contain an integer number of unit cells each. We choose $\mathrm{C}$ to cover all boundaries of $\Lambda_\mathrm{F}$. A particular boundary region $\mathrm{c} \subset \mathrm{C}$ has charge
\begin{equation}
\label{eq:bdrychargedef}
Q_\mathrm{c} = \sum_{\bs{v} \in \mathrm{c}} \sum_\mu \sum_{n \in \mathrm{occ}} \left|\braket{\bs{v} \mu | n}\right|^2,
\end{equation}
where $\ket{n}$ denotes an eigenstate of $H_\mathrm{F}$ that is taken out of the occupied subspace $\mathrm{occ}$ bounded by $E_\mathrm{Fermi}$ and we have $\ket{\bs{v} \mu} = c^\dagger_{\bs{v} \mu} \ket{0}$ where $\ket{0}$ denotes the electronic vacuum. Since we only consider regions $\mathrm{c}$ that are related to each other by elements of $\mathcal{G}_\mathrm{F}$, they have necessarily the same charge. Now, note that the charge of the full system is an even integer (given by $|\mathrm{occ}|$), as is the charge of the complement of $\mathrm{C}$, as long as we choose the regions in $\mathrm{C}$ large enough so as to ensure that the eigenstates localized in the complement are pure bulk-like in character and unaffected by the presence of a boundary. This is always possible when the linear extent by which $\mathrm{C}$ penetrates the bulk is much larger than the correlation length set by the bulk gap. We may then view the states contributing to the charge of the complement of $\mathrm{C}$ as states of a complete system of reduced size that has periodic boundary conditions and even integer charge. We thus deduce that $Q_\mathrm{c}$ is quantized in even integer multiples of $1/q$, where $q = |\mathrm{C}|$ denotes the number of elements in $\mathrm{C}$. See Figs.~\ref{fig:cornerchargefrac}~\textbf{a} and~\textbf{b} for an example with threefold rotational symmetry.

We call $Q_\mathrm{c}$ the corner charge since, in a pristine OAL, its fractional part derives from exponentially localized Wannier orbitals that are ``cut through" by corners in the boundary of the system~\cite{miertcorners,EzawaWannier19,benalcazar2018quantization}: The Wannier orbitals in OALs are localized at maximal Wyckoff positions in the unit cell, and have shapes that respect the little group of their Wyckoff position. When a Wyckoff position lies on the boundary of the unit cell, the latter cuts through the respective Wannier orbital. The corner charge $Q_\mathrm{c}$ can then be calculated conveniently and is equal to the volume that all occupied Wannier functions integrate to in $\mathrm{c}$ (where a single Wannier function is normalized to unit volume).

Note that Wannier orbitals which are cut through by edges instead of corners contribute to the TRS polarization~\cite{z2spinpumpfukane,benalcazar2018quantization,MirrorInsulatorOrtix} and thereby correspond to a charge that scales linearly with the extent of the boundary. The corner charge, on the other hand, stays constant as the thermodynamic limit is taken. It is thus well defined only in absence of TRS polarization.

Importantly, not all OALs have a fractional corner charge in all finite geometries. For example, as we will see for the $1a$ OAL discussed in section~\ref{sec:materials}, sometimes there are no symmetry-preserving terminations that cut through Wannier functions (if only entire unit cells are retained), even though the latter are not centered at the atomic positions of the crystal.

Any trivial atomic limit has $Q_\mathrm{c} \in 2\mathbb{Z}$ for any such choice of boundary region: when different unit cells are not coupled to each other, the charge in each unit cell has to be equal to the total charge of the occupied subspace of $H_{\mathrm{F} = \{(0,0)\}}$, which is necessarily an even integer. We may then define \emph{corner charge fractionalization} as occurring in systems for which $Q_\mathrm{c} \mod 2$ is equal to non-zero even integer multiples of $1/q$ (odd integer multiples are forbidden by TRS). Note that in this work we assume all systems with nontrivial corner charge to be given by OALs, which have a representation in terms of exponentially localized Wannier functions~\cite{VanderbiltReview,Bradlyn17}. However, the corner charge formulas we supply in section~\ref{sec:mappingformulae} apply equally well to fragile phases~\cite{AshvinFragile,BarryFragile,KoreanTBG,wieder2018axion,MaiaFragile1,MaiaFragile2,MaiaFragile3}. These can always be adiabatically continued into OALs when other OALs are added. For a calculation of the corner charge in spinful materials, such a ``trivialization" of a fragile phase becomes necessary only in the symmetry class that has $C_4$ rotational symmetry as its sole crystalline component, since this symmetry does not by itself allow for explicit corner charge formulas in terms of the elementary topological invariants we consider.

The classification of corner charge fractionalization in class AII and symmetry group $\mathcal{G}$ is given by the set of inequivalent $Q_\mathrm{c} \mod 2$ that cannot be changed without breaking $\mathcal{G}_\mathrm{F}$ or closing the bulk gap. We present the classification for all layer groups $\mathcal{G}$ in section~\ref{sec:layergroupcornerclassification} of the Appendix.

\subsection{Robustness of the corner charge}
\label{sec:bdrychargerobustness}
We now discuss to what extent symmetry-preserving edge manipulations can change the corner charge $Q_\mathrm{c}$ defined in Eq.~\eqref{eq:bdrychargedef}. We treat an edge manipulation as the introduction of an additional 1D system along the circumference of the finite 2D sample, and ask how the corner charges of the combined system, defined on the appropriately augmented Hilbert space, can differ from those of the original 2D model. Since charges are additive it is enough to determine the possible charges of the 1D system. 
In the following, we take $Q$ to be the total charge of the 1D addition. It is even due to the requirement that we may only add complete and non-anomalous gapped 1D systems with TRS. We then use the remaining crystalline symmetries to derive further constraints on the charges $Q_{\mathrm{c}}$ that the 1D system contributes to a boundary region $\mathrm{c}$.

We note that the point group symmetries in 2D that $\mathcal{G}_\mathrm{F}$ can contain are mirror and $n$-fold rotational symmetries, where $n \in \{2,3,4,6\}$. We first discuss the latter case of $C_n$ rotational symmetries. For spinful systems with TRS, we have $(C_n)^n = -1$. Let $H_{\mathrm{1D}}$ denote a general 1D TRS gapped Hamiltonian defined on a Hilbert space of $L$ lattice sites (with $L/n$ an integer), possibly augmented by orbital degrees of freedom (see also Fig.~\ref{fig:cornerchargefrac}~\textbf{c}). A $C_n$ rotational symmetry
\begin{equation}
C_n H_{\mathrm{1D}} C_n^\dagger = H_{\mathrm{1D}},
\end{equation}
implies that we can choose the order of regions $\mathrm{c}_i \in \mathrm{C}$ (which in combination cover all of the $L$ sites of the 1D system) such that in real space the symmetry effects $\mathrm{c}_i \rightarrow \mathrm{c}_{i+1 \mod n}$, that is, a translation by $L/n$ sites.
Now, due to $(C_n)^n = -1$, rotations are equivalent to translations around a 1D circle that encloses a $\pi$-flux. Let $t$ be the operator for translations by a single site, i.e., it shifts site $r \in \{1,\dots,L\}$ of the 1D lattice to site $r+1 \mod L$. It is not a symmetry of $H_{\mathrm{1D}}$, however, we can obtain a $t$-symmetric Hamiltonian (on a ring enclosing a $\pi$-flux) by adding up $L/n$ copies of $H_{\mathrm{1D}}$ that are subsequently shifted by one lattice site, to arrive at
\begin{equation}
H_{\mathrm{1D}}^{\mathrm{TRN}} = H_{\mathrm{1D}} \oplus t H_{\mathrm{1D}} t^\dagger \oplus \dots \oplus t^{L/n-1} H_{\mathrm{1D}} (t^\dagger)^{L/n-1},
\end{equation}
which acts on an $L/n$-fold enlarged Hilbert space. The occupied subspace of $H_{\mathrm{1D}}^{\mathrm{TRN}}$ has a total charge of $Q L/n$ and enjoys a translational symmetry that corresponds to $L$ repeated unit cells, with twisted boundary conditions so as to accommodate the $\pi$-flux. It is gapped and has TRS just as $H_{\mathrm{1D}}$, and its charge thus necessarily corresponds to an even integer number of filled Bloch bands, which each hold $L$ states. We conclude that its charge per unit cell $Q/n$ is an even integer. Returning our attention to $H_{\mathrm{1D}}$, since all boundaries $\mathrm{c}$ carry the same charge, this is exactly the corner charge $Q_{\mathrm{c}} = Q/n$. Thus in the case of $C_n$-symmetries there is no 1D addition that can trivialize the fractional corner charges of a 2D OAL.

Next, we turn to mirror symmetries, which for spinful systems satisfy $M^2 = -1$. In the case of two reflections, say $M_x$ and $M_y$, we also have a two-fold rotation symmetry $C_2 = M_x M_y$, which by the argument above allows us to conclude that all corner charges $Q_{\mathrm{c}}$ contributed by any gapped and TRS 1D addition are necessarily even (note that the minimal nontrivial boundary decomposition has $q = 2$). When there is only a single mirror symmetry, we cannot argue along these lines, since it does not act on the 1D real space as a translation. In fact, it ``translates'' different sites along the 1D chain by different amounts. Hence, here the symmetry constraint on $Q_{\mathrm{c}}$ is the same as that for the 2D bulk, namely that $Q_{\mathrm{c}_i} \in \mathbb{Z}$, $i=1,2$, (compare this to the $Q_{\mathrm{c}_i} \in 2\mathbb{Z}$ we obtain for $C_2$ symmetry) and a fractional charge of $1 \mod 2$ can be trivialized.

Finally, we note that in the case where we have $C_3$ symmetry as well as 3D inversion symmetry $\mathcal{I}$ (which is the same as $C_2 M_z$ symmetry), we can define an effective $1/6$ translation by $t_{1/6} = \mathcal{I} C_3^2$ which allows to argue that patches $\mathrm{c}$ of size $1/6$ of the linear extent of the full 1D system have even integer charge. This is important for the robustness of the $Q_\mathrm{c} = 1/6$ corner charges of this symmetry class.

Since any finite-size geometry breaks the remaining non-symmorphic symmetries a system might have, we do not need to consider their effect on charge fractionalization. We conclude that quantized corner charges can be changed by 1D edge manipulations only in the case of a single mirror symmetry.

\section{Identification of obstructed atomic limits}
\label{sec: topo invariants}
In this section we give a prescription for obtaining the corner charge of the occupied subspace of a bulk model represented by a Bloch Hamiltonian $\mathcal{H}(\bs{k})$ [the Fourier transform of the translationally invariant tight-binding model given in Eq.~\eqref{eq:tbham}], assuming that its occupied subspace realizes an OAL. We take a Wannier center point of view: in particular, we define an OAL by the way it is built up from exponentially localized and symmetric Wannier functions\cite{VanderbiltReview,Bradlyn17}.

As shown in section~\ref{sec:bdrychargerobustness}, mirror symmetries can protect fractional corner charges only when they are combined to yield a twofold rotational symmetry. The protecting symmetries we consider are therefore $C_n$ rotations, with or without an additional 3D inversion symmetry $\mathcal{I}$. The inclusion of $\mathcal{I}$ symmetry allows us to extend our discussion to the experimentally relevant case of 2D honeycomb monolayers with nonzero buckling, and our classification (given in the Appendix) to the 80 layer groups instead of the 17 wallpaper groups. We note that inversion effectively replaces $C_2$ in its role of enforcing a $Q_\mathrm{c} = 0,1\mod 2$ quantization of the corner charge, but due to $\mathcal{I}^2 = +1$ [whereas $(C_2)^2 = -1$] allows for symmetry indicator invariants. Furthermore, in the case of $C_4$ symmetry, we find that we require an additional inversion symmetry in order to be able to read off the corner charge from the available topological invariants. Inversion symmetry is however, unlike $C_4$, not necessary for the topological robustness of the corner charge.

We first establish which topological invariants can be defined for each point group in~\ref{sec:topologicalindices}.
We then go on to calculate these indices for the elementary band representations of each symmetry class in~\ref{sec:EBRdecomposition}.
Finally, in~\ref{sec:mappingformulae} we present formulas that allow for a determination of the corner charge in all symmetry classes except for the one that as its only crystalline component has $C_4$ rotational symmetry.

A list of all possible corner charges for the various layer groups is given in Table~\ref{tab:relevantlayergroups} of the Appendix. The remaining symmetry operations a layer group may contain in addition to the ones listed in section~\ref{sec:mappingformulae} are either irrelevant for finite-size corner terminations since they involve translations (as in the case of non-symmorphic symmetries), or merely impose constraints for the shape of the finite-size termination, without affecting the corner charge quantization itself (such as mirror symmetries).

\begin{figure}
\centering
\includegraphics[width=0.48\textwidth]{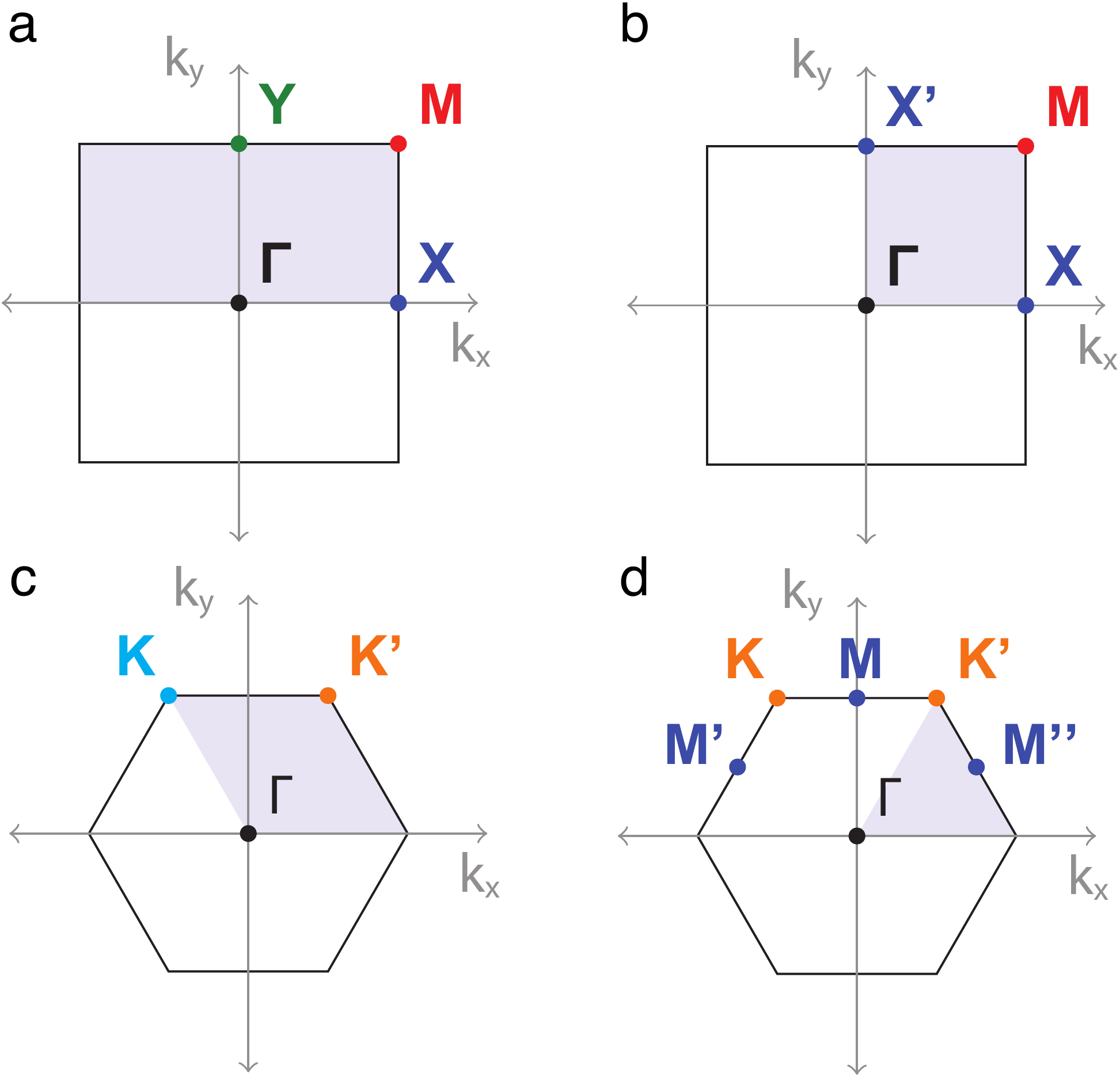}
\caption{Brillouin zones of crystals with $C_2$, $C_4$, $C_3$, and $C_6$ symmetries and their rotation invariant points. In $C_2$-symmetric systems there are three 2-fold HSPs: $\bs{X}$, $\bs{Y}$, and $\bs{M}$. In $C_4$-symmetric systems there are two 2-fold HSPs: $\bs{X}$ and $\bs{X'}$, and one 4-fold HSP: $\bs{M}$. In $C_3$-symmetric systems there are only two 3-fold HSPs: $\bs{K}$ and $\bs{K'}$. Finally, in $C_6$-symmetric systems there are three 2-fold HSPs: $\bs{M}$, $\bs{M'}$, and $\bs{M''}$, as well as two 3-fold HSPs: $\bs{K}$ and $\bs{K'}$.}
\label{fig:BZ}
\end{figure}

\begin{figure}
\centering
\includegraphics[width=0.48\textwidth]{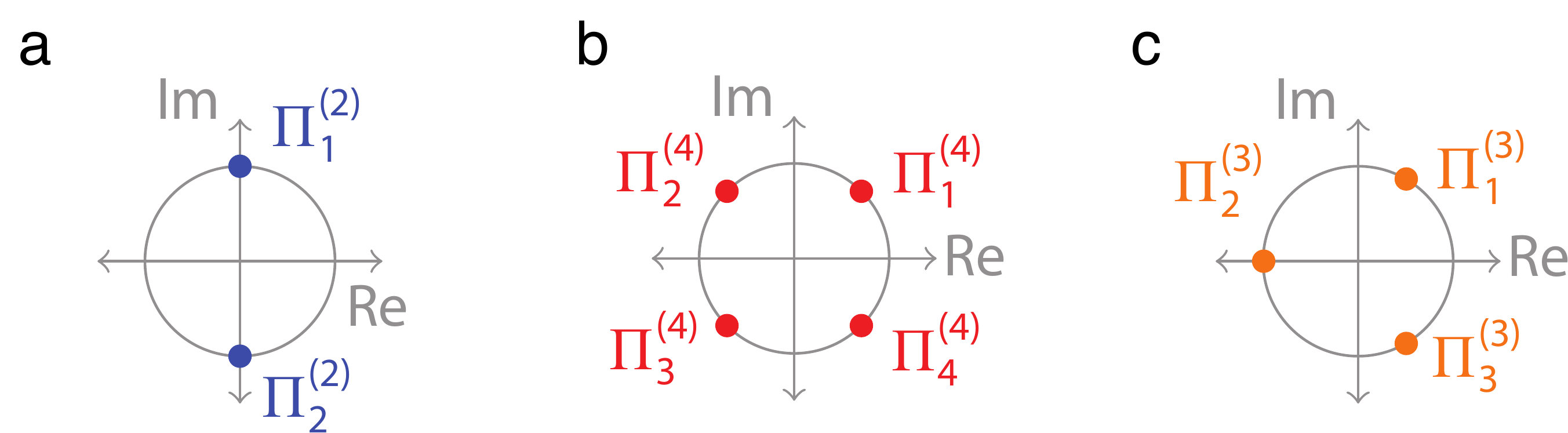}
\caption{Sets of allowed eigenvalues for spinful rotational symmetries. \textbf{a} $C_2$ symmetry. \textbf{b} $C_4$ symmetry. \textbf{c} $C_3$ symmetry. The possible eigenvalues of $C_6$ symmetry are not shown, since they do not allow for the definition of symmetry indicators (there is at most one $C_6$-symmetric point in any two-dimensional Brillouin zone).}
\label{fig:eigvalsets}
\end{figure}

\subsection{Bulk topological indices}
\label{sec:topologicalindices}
To identify different EBRs, we employ a combination of symmetry indicator~\cite{Hughes11,fang2012,Slager12,benalcazar2014,Slager17,ashvin230indicators,Bradlyn17,SongZhang17,Khalaf17,benalcazar2018quantization} and Wilson loop~\cite{WilczekZee,BerryOriginal,ZakPhase1,ZakPhase2,Slager12,Alexandradinata14,Slager17,BarryFragile} topological invariants. These can be evaluated from the crystal's Bloch Hamiltonian $\mathcal{H}(\bs{k})$ and so do not require a real-space calculation to be performed. The main ingredient for both kinds of invariants is the bundle of occupied Bloch states $\ket{u_{m}(\bs{k})}$, $m = 1 \dots N$. [$\bs{k}$ is an element of the first Brillouin zone (BZ) of the crystal.]

Given a unitary crystal symmetry $\mathcal{S}$ that is realized on the Bloch Hamiltonian as $\mathcal{S} \mathcal{H}(\bs{k}) \mathcal{S}^\dagger = \mathcal{H}(\mathsf{S} \bs{k})$ and acts on the momenta as $\bs{k} \rightarrow \mathsf{S} \bs{k}$, we can calculate its corresponding symmetry indicator topological invariants from the eigenvalues of the matrices
\begin{equation}
S_{mn} = \bra{u_m(\bs{\bar{k}})} \mathcal{S} \ket{u_n(\bs{\bar{k}})},
\end{equation}
where $m,n$ run over the occupied subspace only and $\bs{\bar{k}} = \mathsf{S} \bs{\bar{k}}$ are high-symmetry points (HSPs) of the Brillouin zone that are left invariant by the symmetry $\mathcal{S}$ (see also Fig.~\ref{fig:BZ}). An $n$-fold symmetry acting on spinful fermions satisfies $\mathcal{S}^n = \pm1$ (positive sign for 3D inversion, negative sign for 2D mirror and rotational symmetries), this, together with TRS, imposes constraints on the possible eigenvalues of $S_{mn}$ and allows for the definition of topological invariants that capture the different symmetry representations of the occupied bands across the BZ. A trivial OAL, being deformable to a momentum-independent Hamiltonian, will have the same representation across HSPs that are invariant under the same symmetry and hence will have trivial symmetry indicator invariants. Nonzero symmetry indicator invariants, on the other hand, indicate that the bands adopt different representations of the symmetry across the BZ and correspond to nontrivial OALs.

The Wilson loop (along a closed, non-contractible path $\gamma$ in the BZ that starts and ends at the momentum $\bs{k}^*$) is an operator on the filled band subspace of $\mathcal{H}(\bs{k})$ defined as
\begin{equation}
W_\gamma = \prod_{\bs{k}}^{\gamma} P(\bs{k}),
\end{equation}
where $P(\bs{k})= \sum_{m \in \mathrm{occ}} \ket{u_m(\bs{k})}\bra{u_m(\bs{k})}$ is the projector onto the subspace of filled bands at momentum $\bs{k}$. Note that we choose a gauge where $\mathcal{H}(\bs{k}) = \mathcal{H}(\bs{k}+\bs{G})$ for a reciprocal lattice vector $\bs{G}$ and the product is path-ordered along $\gamma$. The Wilson loop operator satisfies $W_\gamma W_\gamma^\dagger = P(\bs{k}^*)$ and so, since any projector satisfies $[P(\bs{k})]^2 = P(\bs{k})$, its eigenvalues are either zero or of the form $e^{\mathrm{i} \theta_\alpha^\gamma}$, $\alpha = 1 \dots N$. In the following, we refer to the set of $\{\theta_\alpha^\gamma\}_{\alpha = 1 \dots N}$ as the Wilson loop spectrum, suppressing the zero eigenvalues.

The anti-unitary TRS $\mathcal{T}$ acts on the Bloch Hamiltonian as $\mathcal{T} \mathcal{H}(\bs{k}) \mathcal{T}^{-1} = \mathcal{H}(-\bs{k})$. For the projectors this implies $\mathcal{T} P(\bs{k}) \mathcal{T}^{-1} = P(-\bs{k})$. When $\gamma$ is mapped onto itself by TRS, and its starting point satisfies $\bs{k}^* = -\bs{k}^*$ up to a reciprocal lattice vector, we then have 
\begin{equation}
\label{eq:TRSonWilson}
\mathcal{T} W_\gamma \mathcal{T}^\dagger = \prod_{\bs{k}}^{\gamma} P(-\bs{k}) = W_\gamma^\dagger.
\end{equation}
Due to $\mathcal{T}$ being anti-unitary and $\mathcal{T}^2 = -1$, this implies a Kramers degeneracy of the Wilson loop spectrum, i.e., every $\theta_\alpha^\gamma$ is (at least) two-fold degenerate when $\gamma$ is mapped onto itself by time reversal.

Now, if there is a crystal symmetry $\mathcal{S}$ that reverses the direction of $\gamma$ and leaves the starting point invariant so that $\bs{k}^* = \mathsf{S} \bs{k}^*$ up to a reciprocal lattice vector, we have 
\begin{equation}
\label{eq:reflectiononWilson}
\mathcal{S} W_\gamma \mathcal{S}^\dagger =  \prod_{\bs{k}}^{\gamma} \left[\mathcal{S} P(\bs{k}) \mathcal{S}^\dagger \right]
= W_\gamma^\dagger.
\end{equation}
Since $\mathcal{S}$ is unitary, the Wilson loop is unitarily equivalent to its complex conjugate and so its eigenvalues come in complex conjugated pairs. This implies a symmetry of the Wilson loop spectrum around $\theta=0$, for every $\theta_\alpha^\gamma$ there is a corresponding $-\theta_\alpha^\gamma \mod 2\pi$.

We may furthermore employ nested Wilson loops~\cite{BABHughesBenalcazar17,Benalcazar17}. Let $W_i (k_j)$, $i \neq j$, denote the Wilson loop along the noncontractible loop $\gamma: (k_i = 0, k_j) \rightarrow (k_i = 2\pi, k_j)$, where ($k_i$, $k_j$) labels a point in the two-dimensional BZ in some basis (chosen such that $k_{i,j} = 0$ and $k_{i,j} = 2\pi$ are related by reciprocal lattice vectors). Consider the Wilson loop Hamiltonian $H_{W_i} (k_j)$, defined by
\begin{equation}
\left[e^{\mathrm{i} H_{W_i}(k_j)} \right]_{mn}= \bra{u_m(k_i=0,k_j)} W_i (k_j) \ket{u_n(k_i=0,k_j)}.
\end{equation}
Equations~\eqref{eq:TRSonWilson} and~\eqref{eq:reflectiononWilson} then imply
\begin{equation}
\begin{aligned}
\mathcal{T}_{k_j} H_{W_i} (k_j) \mathcal{T}_{k_j}^\dagger &= H_{W_i} (-k_j), \\
\mathcal{S}_{k_j} H_{W_i} (k_j) \mathcal{S}_{k_j}^\dagger &= - H_{W_i} (\mathsf{S} k_j),
\end{aligned}
\label{eq:nestedWilsonprops}
\end{equation}
where we defined
\begin{equation}
\begin{aligned}
\left(\mathcal{T}_{k_j} \right)_{mn} &= \bra{u_m(-k_j)} \mathcal{T} \ket{u_n(k_j)}, \\
\left(\mathcal{S}_{k_j} \right)_{mn} &= \bra{u_m(\mathsf{S}k_j)} \mathcal{S} \ket{u_n(k_j)}.
\end{aligned}
\end{equation}
We see that $\mathcal{T}$ implies a TRS of the Wilson loop Hamiltonian, whereas $\mathcal{S}$ implies a particle-hole symmetry. These properties are needed for the definition of quantized topological invariants of the \emph{nested Wilson loop}: We define $W_i^\mathrm{b}$ as the Wilson loop calculated from a gapped set eigenstates $\mathrm{b}$ of $H_{W_i} (k_j)$ along a closed, non-contractible path $k_j: 0 \rightarrow 2\pi$ in the reduced BZ.

We differentiate between three kinds of nested Wilson loops that differ by the choice of the set of eigenstates $\mathrm{b}$: 1) The nested loop $W_i^0$, which is calculated from the two bands in the spectrum of $H_{W_i} (k_j)$ that at $k_j = 0, \pi$ have a degeneracy pinned to the Wilson eigenvalue $0$ [note that any such degeneracy at $k_j = 0$ implies one at $k_j = \pi$ and vice versa due to the absence of Wannier center flow in (obstructed) atomic limits]. 2) The nested Wilson loop $W_i^{\lambda}$, calculated for the upper \emph{or} lower half of the bands in the spectrum of $H_{W_i} (k_j)$ that are \emph{not} pinned at $k_j = 0, \pi$ to a Wilson eigenvalue $0,\pi$ (that is, half of the freely dangling Wilson bands, which by the particle hole symmetry come in pairs). 
3) The nested loop $W_i^{\pi}$, which is calculated from the two bands in the spectrum of $H_{W_i} (k_j)$ that at $k_j = 0, \pi$ have a degeneracy pinned to the Wilson eigenvalue $\pi$.

The nested Wilson loops of type (1) and (3) cannot be trivialized by transformations that preserve $\mathcal{S}$ and $\mathcal{T}$ and are adiabatic with respect to the bulk gap. The reason is that the invariants calculated from these loops are equal to the partial polarizations of Wilson bands pinned to eigenvalues $0$ or $\pi$ by $\mathcal{S}$ at the transverse momenta $k_j = 0, \pi$. Wilson gap closings that preserve the energy gap can only occur in pairs (due to $\mathcal{S}$) at intermediate transverse momenta $k_j, -k_j$. It is rigorously shown in Appendix A of Ref.~\onlinecite{Sander2DTCI2019} that these gap closings together always contribute integer multiples of $2\pi$ to the nested partial polarization, and therefore cannot trivialize Wilson loops of type (1) and (3). In this work, we do not consider invariants derived from Wilson loops of type (2).

We will now list the topological invariants that can be defined for a given point group. We find that often the inclusion of $\mathcal{I}$ symmetry allows for the replacement of Wilson-loop invariants by symmetry indicator invariants.
For the discussion of symmetry indicators, we make use of the definitions and derivations presented in sections~\ref{sec:rotsymwladimir}-\ref{sec:trscnstrnswladimir} of the Appendix.

Note that in the following, and as motivated at the beginning of section~\ref{sec:bdrychquant}, we explicitly exclude invariants that characterize topological insulators because they are necessarily gapless along the edges in a 2D geometry with open boundary conditions, and so do not allow for stable quantized corner charges. In addition to removing some invariants from our analysis altogether, this imposes constraints on Wilson loops.

We emphasize that our list of invariants may not be exhaustive. As noted in Ref.~\onlinecite{BarryFragile}, it is in general difficult to identify all possibly nontrivial Wilson loop invariants. In this work we only treat ``straight" (nested) Wilson loops, which (given a starting point) go around one of the two inequivalent noncontractible loops of the Brillouin zone torus.

\subsubsection{$C_2$ symmetry}

\paragraph{Symmetry indicator invariants}
The BZ has four high-symmetry points (HSPs, defined in section~\ref{sec:invptswladimir} of the Appendix), see also Fig.~\ref{fig:BZ}~\textbf{a}. All the points are invariant under $C_2$. Thus, they all have $C_2$ eigenvalues ${+i,-i}$ (see Fig.~\ref{fig:eigvalsets}~\textbf{a}). However, since all the HSPs are also time-reversal invariant momenta (TRIMs), the eigenvalues have to come in complex-conjugate pairs, leading to a single available 2D irreducible representation. Therefore, the $C_2$ eigenvalues on their own do not afford a topological distinction and there are no symmetry indicator invariants.

\paragraph{Wilson-loop invariants}
For every closed high-symmetry line $\gamma$ (which connects two HSPs) of the 2D BZ that is left invariant by $C_2$, we can define a Wilson loop that is TRS and $C_2$ symmetric. Due to Eqs.~\eqref{eq:TRSonWilson} and ~\eqref{eq:reflectiononWilson} the parities of the numbers of $\theta_\alpha^\gamma = 0$ and $\theta_\alpha^\gamma = \pi$ eigenvalues in its spectrum cannot be changed under adiabatic deformations of $\mathcal{H}(\bs{k})$: adiabatic perturbations of the Hamiltonian at most move particle-hole related Kramers pairs of eigenvalues in and out of $0$ and $\pi$, this does not change the total parity. 

A topological invariant of $W_\gamma$ with spectrum $\{\theta_\alpha^\gamma\}_{\alpha = 1 \dots N}$ is therefore given by
\begin{equation}
\label{eq:wilsonloopwithreflectioninvariant}
\nu_{\gamma} = -\frac{i}{\pi} \log \left( \prod_{\alpha=1,3,\dots,N-1} e^{\mathrm{i} \theta_\alpha^\gamma} \right) \mod 2,
\end{equation}
where the product is taken over only one eigenvalue of each Wilson loop Kramers pair. We call $\nu_{\gamma} = 0$ trivial and $\nu_{\gamma} = 1$ nontrivial. This invariant is equivalent to the TRS polarization~\cite{MirrorInsulatorOrtix} and counts the parity number of Wilson loop pairs of eigenvalues equal to $\pi$. We also define
\begin{equation}
\label{eq:wilsonloopwithreflectioninvariant2}
\mu_{\gamma} = -\frac{i}{\pi} \log \left( \prod_{\alpha=1,3,\dots,N-1} e^{\mathrm{i} (\pi-\theta_\alpha^\gamma)} \right) \mod 2,
\end{equation}
which counts the number of Wilson loop pairs of eigenvalues equal to $0$. The invariants $\nu_{\gamma}$ and $\mu_{\gamma}$ are not independent when the total number of bands is fixed. They obey
\begin{equation}
\mu_{\gamma} = \nu_{\gamma} + \frac{N}{2} \mod 2.
\end{equation}
Therefore, we drop $\mu_\gamma$ as it provides redundant topological information. 
In the following, we will consider Wilson loops that go through high-symmetry points in the 2D BZ. We denote by $\nu_{\bs{A B}}$ the loop that goes from point $\bs{A}$ to point $\bs{B}$ and then back to $\bs{A}$ via the shortest non-contractible loop around the BZ torus.

There are in total four TRIMs and three topologically inequivalent straight and $C_2$-symmetric Wilson loops. This can be seen by noting that, holding one of the four $C_2$-symmetric momenta fixed as a starting point, there are two incontractible loops around the Brillouin zone torus (which necessarily go through one other $C_2$-symmetric momentum). Keeping in mind that path-reversed Wilson loops are not independent [as per Eq.~\eqref{eq:reflectiononWilson}], this naively yields the set of Wilson loop invariants $\{\nu_{\bs{\Gamma X}}, \nu_{\bs{\Gamma Y}}, \nu_{\bs{X M}}, \nu_{\bs{Y M}}\}$. We note however that the path denoted by $\bs{Y \Gamma} - \bs{\Gamma X} - \bs{XM}$ is topologically equivalent to the path denoted by $\bs{Y M}$ and so we have
\begin{equation}
\nu_{\bs{Y M}} = \nu_{\bs{\Gamma Y}} \nu_{\bs{\Gamma X}} \nu_{\bs{X M}}.
\end{equation}
The remaining invariants are further constrained due to the requirement that the $\mathbb{Z}_2$ TI invariant $\Delta_{\mathrm{TI}}$ vanishes: we have that
\begin{equation}
\begin{aligned}
\Delta_{\mathrm{TI}} & = \nu_{\bs{\Gamma X}} + \nu_{\bs{Y M}} \mod 2 \\
& = \nu_{\bs{\Gamma Y}} + \nu_{\bs{X M}} \mod 2.
\end{aligned}
\end{equation}
We are left with two Wilson loop invariants.

Similarly, we may define the quantized invariants $\nu_{x,y}^{0,\pi}$ and $\mu_{x,y}^{0,\pi}$ from the nested Wilson loops $W_{x,y}^{0,\pi}$, since these are calculated for particle-hole symmetric sets of bands~\cite{BenMoTe2,wieder2018axion} [in contrast to $W_{x,y}^{\lambda}$, which does not satisfy Eq.~\eqref{eq:nestedWilsonprops}]: the anti-commutativity with the Wilson loop Hamiltonian that distinguishes particle-hole symmetry from a reflection symmetry is irrelevant from the point of view of the nested Wilson loop, as long as the latter is defined via a projector onto a particle-hole symmetric set of bands. We may therefore define $\nu_{x,y}^{0,\pi}$ and $\mu_{x,y}^{0,\pi}$ just as in Eqs.~\eqref{eq:wilsonloopwithreflectioninvariant} and~\eqref{eq:wilsonloopwithreflectioninvariant2}, where $\theta_\alpha^\gamma$ this time refers to the spectrum of the nested Wilson loop. As before, we drop the $\mu$ invariants since they are not independent when the number of occupied Wilson bands is held fixed. Taking into account the constraints~\cite{Sander2DTCI2019} 
\begin{equation}
\begin{aligned}
\nu_{x}^{0} + \nu_{x}^{\pi} \mod 2 &= \nu_{\bs{\Gamma Y}}, \\
\nu_{y}^{0} + \nu_{y}^{\pi} \mod 2 &= \nu_{\bs{\Gamma X}}, \\
\nu_{x}^{\pi} &= \nu_{y}^{\pi},
\end{aligned}
\end{equation}
reduces the number of independent invariants to three. The third equation can be seen in the following way: $\nu_{x}^{\pi}$ is nonzero if and only if the occupied subspace hosts an odd number of Wannier Kramers pairs whose centers are shifted by $1/2$ in both $x$ and $y$ direction (taking the lattice constant to be $1$) with respect to the center of the unit cell, i.e., if there is an odd number of Kramers pairs at Wyckoff position $1b$ of the crystal (see also Table~\ref{tab:EBRc2}). This Wyckoff position stays unchanged when exchanging $x$ and $y$, we therefore obtain that $\nu_{x}^{\pi}$ is nonzero if and only if $\nu_{y}^{\pi}$ is nonzero. Note that the corresponding statement does not hold for $\nu_{x}^{0}$ and $\nu_{y}^{0}$, since these indicate Wannier Kramers pairs at the $1c$ and $1d$ Wyckoff positions, respectively.

We therefore choose the classification to be given by
\begin{align}
\label{eq:c2classificationWithoutInversion}
\chi^{(2)} = \{\nu_{\bs{\Gamma X}},\nu_{\bs{\Gamma Y}}, \nu_{x}^{\pi}\}.
\end{align}

\paragraph{With inversion symmetry}
$C_2 + \mathcal{I}$ symmetry is equivalent to $\mathcal{I}$ symmetry for all our purposes.
Inversion symmetry allows us to replace the $\mathbb{Z}_2$ valued Wilson-loop invariants by $2\mathbb{Z}$-valued symmetry indicators.
The BZ has the $\mathcal{I}$-invariant points $\bs{\Gamma}$, $\bs{X}$, $\bs{Y}$ and $\bs{M}$, which support the six inversion eigenvalue invariants
\begin{align}
[X^\mathcal{I}_i]&= \#X^\mathcal{I}_i - \#\Gamma^{\mathcal{I}}_i, \nonumber\\
[Y^\mathcal{I}_i]&= \#Y^\mathcal{I}_i - \#\Gamma^{\mathcal{I}}_i, \nonumber\\
[M^\mathcal{I}_i]&= \#M^\mathcal{I}_i - \#\Gamma^{\mathcal{I}}_i,
\end{align}
where $\#X^\mathcal{I}_i$ ($\#\Gamma^{\mathcal{I}}_i$) is the number of occupied states with inversion eigenvalue $X^\mathcal{I}_i$ ($\Gamma^{\mathcal{I}}_i$), and $X^\mathcal{I}_{i=1,2}$, $\Gamma^{\mathcal{I}}_{i=1,2}=\{1,-1\}$, and similarly for $\bs{Y}$ and $\bs{M}$.
Due to the fixed number of occupied bands, we have the constraints
\begin{align}
\label{eq:constraintsoninversioneigvals}
[X^\mathcal{I}_1]+[X^\mathcal{I}_2]&=0, \nonumber\\
[Y^\mathcal{I}_1]+[Y^\mathcal{I}_2]&=0, \nonumber\\
[M^\mathcal{I}_1]+[M^\mathcal{I}_2]&=0.
\end{align}
The three remaining invariants completely fix~\cite{Alexandradinata14} the Wilson loops in Eq.~\eqref{eq:c2classificationWithoutInversion}. Due to TRS they are necessarily even integers. We retain the classification
\begin{align}
\label{eq:c2classificationWithInversion}
\chi^{(2)}_\mathcal{I} = \{[X^\mathcal{I}_2],[Y^\mathcal{I}_2],[M^\mathcal{I}_2]\}.
\end{align}

\subsubsection{$C_3$ symmetry}

\paragraph{Symmetry indicator invariants}
The BZ only has the $C_3$-invariant points $\bs{K}$ and $\bs{K'}$, see also Fig.~\ref{fig:BZ}~\textbf{c}.
Now, we discuss the invariants that compare the representations at the $\bs{K}$ ($\bs{K'}$) and $\bs{\Gamma}$ points of the BZ,
\begin{align}
\label{eq:C3spinfulindicators}
[K^{(3)}_i]= \#K^{(3)}_i - \#\Gamma^{(3)}_i,
\end{align}
where $K^{(3)}_{i=1,2,3}$, $\Gamma^{(3)}_{i=1,2,3}=\{e^{i \pi/3},-1,e^{-i \pi/3}\}$, and similarly for $\bs{K'}$ (see Fig.~\ref{fig:eigvalsets}~\textbf{c}). Unlike $\bs{M}$, the HSP $\bs{K}$ is \emph{not} a TRIM. Instead, TRS relates $\bs{K}$ and $\bs{K'}$. TRS imposes the constraints,
\begin{align}
[K^{(3)}_1] = [K'^{(3)}_3],\nonumber\\
[K^{(3)}_2] = [K'^{(3)}_2],\nonumber\\
[K^{(3)}_3] = [K'^{(3)}_1].
\label{eq:C3_constraints}
\end{align}
The six invariants are subject to the constraints \eqref{eq:C3_constraints} along with
\begin{align}
\label{eq:secondsetofC3indicatorconstraints}
[K^{(3)}_1]+[K^{(3)}_2]+[K^{(3)}_3]=0,\nonumber\\
[K'^{(3)}_1]+[K'^{(3)}_2]+[K'^{(3)}_3]=0.
\end{align}
due to the constant number of occupied states across the BZ.
The symmetry-indicated part of the classification is given by the two invariants
\begin{align}
\chi^{(3)}=\{[K^{(3)}_1],[K^{(3)}_2]\}.
\end{align}

\paragraph{Wilson-loop invariants}
There are no Wilson loop invariants in this class due to the lack of a twofold symmetry.

\paragraph{With inversion symmetry}
Inversion symmetry implies $[K^{(3)}_i] = [K'^{(3)}_i]$, $i = 1,2,3$. We therefore drop $[K^{(3)}_1]$ from the list of independent invariants.
The BZ has the $\mathcal{I}$-invariant points $\bs{M}$, $\bs{M'}$ and $\bs{M''}$, which support the invariants
\begin{align}
[M^\mathcal{I}_i]= \#M^\mathcal{I}_i - \#\Gamma^{\mathcal{I}}_i,
\end{align}
where $M^\mathcal{I}_{i=1,2}$, $\Gamma^{\mathcal{I}}_{i=1,2}=\{1,-1\}$, and similarly for $\bs{M'}$ and $\bs{M''}$. TRS implies that the states belonging to a Kramers pair have equal inversion eigenvalue. $C_3$ imposes the constraints
\begin{align}
[M^\mathcal{I}_1] = [M'^{\mathcal{I}}_1] = [M''^{\mathcal{I}}_1],\nonumber\\
[M^\mathcal{I}_2] = [M'^{\mathcal{I}}_2] = [M''^{\mathcal{I}}_2].
\label{eq:C3_constraintsForInversion}
\end{align}
In addition we have
\begin{equation}
[M^\mathcal{I}_1]+[M^\mathcal{I}_2]=0.
\end{equation}
We retain $[M^\mathcal{I}_2]$ as the invariant that determines the classification, in addition to the $C_3$ invariant $[K^{(3)}_2]$:
\begin{align}
\chi^{(3)}_\mathcal{I}=\{[M^\mathcal{I}_2],[K^{(3)}_2]\}.
\end{align}

\subsubsection{$C_4$ symmetry}

\paragraph{Symmetry indicator invariants}
The BZ has four HSPs (Fig.~\ref{fig:BZ}~\textbf{b}). Two of them are invariant under $C_2$ and give rise to trivial indicators due to time reversal symmetry. 
We can then only build indices that compare the $C_4$ symmetry representations at $\bs{M}$ with those at $\bs{\Gamma}$ as follows:
\begin{align}
[M_i^{(4)}] &= \#M_i^{(4)} - \#\Gamma^{(4)}_i, \quad i \in \{1,2,3,4\}, \nonumber \\
\end{align}
where the eigenvalues are taken from the set $M^{(4)}_{i=1,2,3,4}, \Gamma^{(4)}_{i=1,2,3,4}=\{e^{i \pi/4}, e^{i 3\pi/4}, e^{-i 3\pi/4}, e^{-i \pi/4}\}$, respectively (see Fig.~\ref{fig:eigvalsets}~\textbf{b}).
Since all the HSPs are also TRIMs, the rotation eigenvalues have to come in complex-conjugate pairs. Therefore, we have the constraints on the invariants
\begin{align}
[M_1^{(4)}] &= [M_4^{(4)}],\nonumber\\
[M_2^{(4)}] &= [M_3^{(4)}].
\label{eq:ConstraintsC4_1}
\end{align}
Since the number of occupied states is constant across the BZ, we have that $\sum_i \# M_i^{(4)} = \sum_i \# \Gamma^{(4)}_i$, or
\begin{align}
[M_1^{(4)}]+[M_2^{(4)}]+[M_3^{(4)}]+[M_4^{(4)}]=0.
\label{eq:ConstraintsC4_2}
\end{align}
With the constraints in \eqref{eq:ConstraintsC4_1} and \eqref{eq:ConstraintsC4_2}, we eliminate the redundant invariants $[M_2^{(4)}]$, $[M_3^{(4)}]$, and $[M_4^{(4)}]$. Thus, the classification due to $C_4$ symmetry has only one symmetry-indicator invariant, $[M_1^{(4)}]$.

\paragraph{Wilson-loop invariants}
$C_4$ symmetry implies having $C_2$ symmetry as well and so we can immediately take over the Wilson loops given in Eq.~\eqref{eq:c2classificationWithoutInversion} as possible invariants, where due to $C_4$ we have $\nu_{\bs{\Gamma X}} = \nu_{\bs{\Gamma Y}}$. 

We conclude that the classification is given by
\begin{align}
\chi^{(4)}=\left\{\nu_{\bs{\Gamma X}}, \nu_{x}^{\pi}, [M_1^{(4)}] \right\}.
\end{align}

\paragraph{With inversion symmetry}
The invariants given in Eq.~\eqref{eq:c2classificationWithInversion} (together with the $C_4$ constraint $[X^\mathcal{I}_2] = [Y^\mathcal{I}_2]$) allow us to replace $\nu_{\bs{\Gamma X}}, \nu_{\bs{Y M}}$. 
We conclude that the classification with inversion symmetry is given by
\begin{align}
\chi^{(4)}_\mathcal{I}=\left\{[X^\mathcal{I}_2],[M^\mathcal{I}_2], [M_1^{(4)}]\right\}.
\end{align}

\subsubsection{$C_6$ symmetry}

\paragraph{Symmetry indicator invariants}
In a $C_6$-symmetric BZ, there are two inequivalent HSPs, $\bs{M}$, which is invariant under $C_2$, and $\bs{K}$, which is invariant under $C_3$ (Fig.~\ref{fig:BZ}~\textbf{d}). All other points are related by rotations, and thus provide redundant representations for the purpose of classification. Furthermore,  $\bs{M}$ is both a HSP and a TRIM. Thus, from the analysis of the previous classifications, no invariants can be derived from its representations. Now, we discuss the invariants that compare the representations at the $\bs{K}$ and $\bs{\Gamma}$ points of the BZ,
\begin{align}
[K^{(3)}_i]= \#K^{(3)}_i - \#\Gamma^{(3)}_i,
\end{align}
where $K^{(3)}_{i=1,2,3}$, $\Gamma^{(3)}_{i=1,2,3}=\{e^{i \pi/3},-1,e^{-i \pi/3}\}$. Unlike $\bs{M}$, the HSP $\bs{K}$ is \emph{not} a TRIM. Instead, TRS relates $\bs{K}$ and $\bs{K'}$. TRS imposes the constraints,
\begin{align}
[K^{(3)}_1] = [K'^{(3)}_3],\nonumber\\
[K^{(3)}_2] = [K'^{(3)}_2],\nonumber\\
[K^{(3)}_3] = [K'^{(3)}_1].
\label{eq:C6_constraints}
\end{align}
But the representations at $\bs{K}$ and $\bs{K'}$ are the same due to $C_6$ symmetry,
\begin{align}
[K^{(3)}_1] = [K'^{(3)}_1],\nonumber\\
[K^{(3)}_2] = [K'^{(3)}_2],\nonumber\\
[K^{(3)}_3] = [K'^{(3)}_3].
\end{align}
The last two sets of constraints leave us with only two non-redundant invariants, $[K^{(3)}_1]$ and $[K^{(3)}_2]$. However, due to the constant number of occupied states, we have $\sum_i \# K^{(3)}_i = \sum_i \# \Gamma^{(3)}_i$ or $2[K^{(3)}_1]+[K^{(3)}_2]=0$, which makes one of these invariants redundant too. We choose the symmetry-indicated part of the classification to be given by $[K^{(3)}_2]$.

\begin{figure}
\centering
\includegraphics[width=0.48\textwidth]{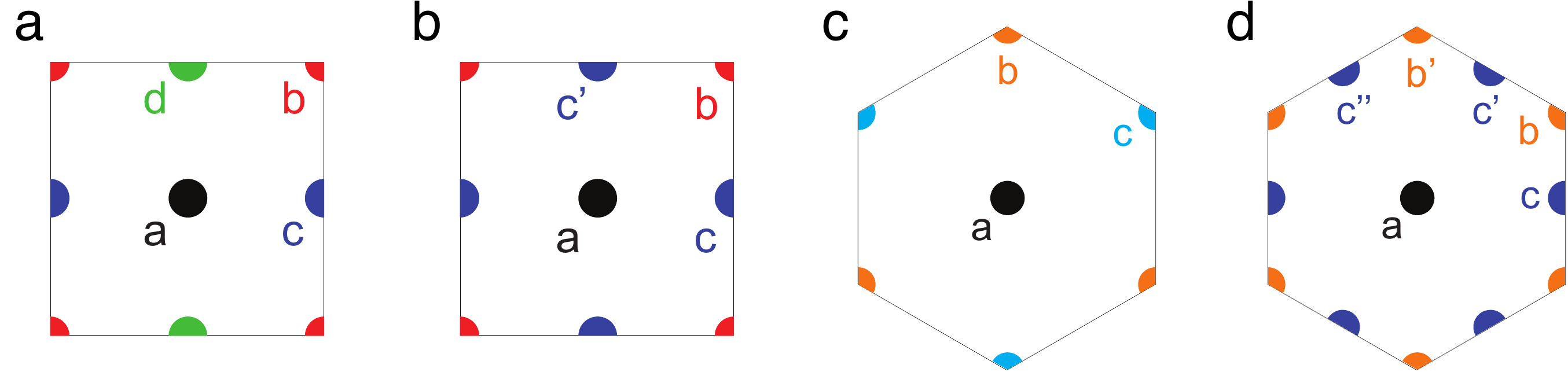}
\caption{Maximal Wyckoff positions for unit cells with rotational symmetry. \textbf{a} $C_2$ symmetry. \textbf{b} $C_4$ symmetry. \textbf{c} $C_3$ symmetry. \textbf{d} $C_6$ symmetry or $C_3 + \mathcal{I}$ symmetry. Boundary charges arise when the Wyckoff positions which Wannier centers are located at are cut through by the crystal termination.}
\label{fig:wyckpositions}
\end{figure}

\paragraph{Wilson-loop invariants}
$C_6$ symmetry implies having $C_2$ symmetry as well and so we can define $\mu_{\bs{\Gamma M}}$ as an invariant due to Eq.~\eqref{eq:wilsonloopwithreflectioninvariant2}. We choose $\mu_{\bs{\Gamma M}}$ here instead of $\nu_{\bs{\Gamma M}}$ since it directly indicates Wannier centers at the $3c$ Wyckoff position (see Fig.~\ref{fig:wyckpositions}~\textbf{d} and Table~\ref{tab:EBRc6}) of the hexagonal unit cell. We do not consider nested Wilson loops in this symmetry class because the corner charge can be completely determined without them.
In conclusion, we have
\begin{align}
\chi^{(6)}=\{\mu_{\bs{\Gamma M}},[K^{(3)}_2]\}.
\end{align}

\paragraph{With inversion symmetry}
The BZ has the $\mathcal{I}$-invariant points $\bs{M}$, $\bs{M'}$ and $\bs{M''}$, which support the invariants
\begin{align}
[M^\mathcal{I}_i]= \#M^\mathcal{I}_i - \#\Gamma^{\mathcal{I}}_i,
\end{align}
where $M^\mathcal{I}_{i=1,2}$, $\Gamma^{\mathcal{I}}_{i=1,2}=\{1,-1\}$, and similarly for $\bs{M'}$ and $\bs{M''}$. TRS implies that the states belonging to a Kramers pair have equal inversion eigenvalue. $C_6$ imposes the constraints
\begin{align}
[M^\mathcal{I}_1] = [M'^{\mathcal{I}}_1] = [M''^{\mathcal{I}}_1],\nonumber\\
[M^\mathcal{I}_2] = [M'^{\mathcal{I}}_2] = [M''^{\mathcal{I}}_2].
\label{eq:C6_constraintsForInversion}
\end{align}
In addition we have
\begin{equation}
[M^\mathcal{I}_1]+[M^\mathcal{I}_2]=0.
\end{equation}
We retain $[M^\mathcal{I}_2]$ as the invariant that determines the classification. Due to $C_6$ symmetry\cite{benalcazar2018quantization} and TRS, $[M^\mathcal{I}_2] \in 4 \mathbb{Z}$. We conclude that
\begin{align}
\chi^{(6)}_\mathcal{I}=\{[M^\mathcal{I}_2],[K^{(3)}_2]\}.
\end{align}

\subsubsection{Summary}
\begin{table}[H]
\centering
\begin{tabular}{|c||c|c|}
\hline
$\mathcal{S}$ & without $\mathcal{I}$ & with $\mathcal{I}$ \\ \hhline{|=#=|=|}
$\mathcal{I}$ & none & $[X^\mathcal{I}_2],[Y^\mathcal{I}_2],[M^\mathcal{I}_2]$ \\ \hline
$C_2$  & $\nu_{\bs{\Gamma X}},\nu_{\bs{\Gamma Y}},\nu_{x}^{\pi}$                      & $[X^\mathcal{I}_2],[Y^\mathcal{I}_2],[M^\mathcal{I}_2]$                      \\ \hline
$C_3$  & $[K^{(3)}_1],[K^{(3)}_2]$                      & $[M^\mathcal{I}_2],[K^{(3)}_2]$                      \\ \hline
$C_4$  & $\nu_{\bs{\Gamma X}}, \nu_{x}^{\pi}, [M_1^{(4)}]$                      & $[X^\mathcal{I}_2],[M^\mathcal{I}_2], [M_1^{(4)}]
$                      \\ \hline
$C_6$  & $\mu_{\bs{\Gamma M}},[K^{(3)}_2]$                      & $[M^\mathcal{I}_2],[K^{(3)}_2]$                      \\ \hline
\end{tabular}
\caption{\label{tab:summaryofindicators} Summary of Wilson loop and symmetry indicator invariants.}
\end{table}

\subsection{Decomposition into EBRs}
\label{sec:EBRdecomposition}
Tables~\ref{tab:EBRc2}-\ref{tab:EBRc6} list the EBRs~\cite{Slager12,Slager17,Bradlyn17,Cano17,Cano17-2,BarryFragile,ZakEBR1,ZakEBR2} supported by systems with $C_n$ rotational symmetry, together with their invariants and corner charges. The minimal block sizes correspond to the multiplicities of the respective Wyckoff positions (multiplied by two to account for spin). If multiple choices for the site-symmetry group~\cite{Bradlyn17} representation at a Wyckoff position $W$ are available, we denote the representation with eigenvalues $e^{\mathrm{i} \alpha}$ as $W|_{\alpha}$.

We show in Sec.~\ref{sec:inductionEBRs} of the Appendix how the symmetry indicator invariants for different EBRs can be derived. The (nested) Wilson loop invariants can be obtained by the mapping of Wilson loop spectra to Wannier centers~\cite{VanderbiltReview}.

\begin{table}[H]
\centering
\begin{tabular}{|c||c|c|c||c|}
\hline
$C_2$ & $\nu_{\bs{\Gamma X}}$ & $\nu_{\bs{\Gamma Y}}$ & $\nu_{x}^{\pi}$ & $Q_\mathrm{c}$ \\ \hhline{|=#=|=|=#=|}
$1a$  & 0                      & 0            & 0          & 0               \\ \hline
$1b$  & 1                      & 1            & 1          & 1               \\ \hline
$1c$  & 1                      & 0            & 0          & 0               \\ \hline
$1d$  & 0                      & 1            & 0          & 0               \\ \hline
\end{tabular}
\caption{\label{tab:EBRc2} EBRs with $C_2$ symmetry induced from the maximal Wyckoff positions listed in the first column (see Fig.~\ref{fig:wyckpositions}~\textbf{a}), and their invariants. All atomic limits can be decomposed into EBRs formed by single Kramers pairs.}
\end{table}

\begin{table}[H]
\centering
\begin{tabular}{|c||c|c|c||c|}
\hline
$C_4$ & $\nu_{\bs{\Gamma X}}$ & $\nu_{x}^{\pi}$  & $[M_1^{(4)}]$ & $Q_\mathrm{c}$ \\ \hhline{|=#=|=|=#=|}
$1a$  & 0          & 0           & 0                       & 0              \\ \hline
$1b|_{\pm \frac{\pi}{4}}$  & 1          & 1           & -1                       & 1/2            \\ \hline
$1b|_{\pm \frac{3\pi}{4}}$  & 1          & 1           & 1                       & 1/2            \\ \hline
$2c$  & 1          & 0           & 0                       & 0                \\ \hline
$1b|_{\pm \frac{\pi}{4}} \oplus  1b|_{\pm \frac{3\pi}{4}}$  & 0          & 0           & 0                       & 1            \\ \hline
\end{tabular}
\caption{\label{tab:EBRc4} EBRs with $C_4$ symmetry induced from the maximal Wyckoff positions listed in the first column (see Fig.~\ref{fig:wyckpositions}~\textbf{b}), and their invariants. All atomic limits can be decomposed into EBRs formed by at most two Kramers pairs. Importantly, the (non-elementary) band representation $1b|_{\pm \frac{\pi}{4}} \oplus  1b|_{\pm \frac{3\pi}{4}}$ has trivial $C_4$ invariants but nonzero corner charge. In systems with $C_4$ as the sole crystalline symmetry, this obstructs a determination of the corner charge in terms of topological invariants.
}
\end{table}

\begin{table}[H]
\centering
\begin{tabular}{|c||c|c||c|}
\hline
$C_3$ & $[K^{(3)}_1]$ & $[K^{(3)}_2]$ & $Q_\mathrm{c}$ \\ \hhline{|=#=|=#=|}
$1a$  & 0                      & 0                      & 0               \\ \hline
$1b|_{\pi}$  & 0                      & -2                      & 2/3               \\ \hline
$1b|_{\pm \frac{\pi}{3}}$  & 0                      & 1                      & 2/3               \\ \hline
$1c|_{\pi}$  & 2                      & -2                      & 0               \\ \hline
$1c|_{\pm \frac{\pi}{3}}$  & -1                      & 1                      & 0               \\ \hline
\end{tabular}
\caption{\label{tab:EBRc3} EBRs with $C_3$ symmetry induced from the maximal Wyckoff positions listed in the first column (see Fig.~\ref{fig:wyckpositions}~\textbf{c}), and their invariants. All atomic limits can be decomposed into EBRs formed by single Kramers pairs.}
\end{table}

\begin{table}[H]
\centering
\begin{tabular}{|c||c|c||c|}
\hline
$C_6$ & $\mu_{\bs{\Gamma M}}$ & $[K^{(3)}_2]$ & $Q_\mathrm{c}$ \\ \hhline{|=#=|=#=|}
$1a$  & 0                      & 0                      & 0               \\ \hline
$2b|_{\pi}$  & 0                      & -4                      & 4/3               \\ \hline
$2b|_{\pm \frac{\pi}{3}}$  & 0                      & 2                      & 4/3               \\ \hline
$3c$  & 1                      & 0                      & 1               \\ \hline
\end{tabular}
\caption{\label{tab:EBRc6} EBRs with $C_6$ symmetry induced from the maximal Wyckoff positions listed in the first column (see Fig.~\ref{fig:wyckpositions}~\textbf{d}), and their invariants. All atomic limits can be decomposed into EBRs formed by at most three Kramers pairs.}
\end{table}

\subsection{Formulas for corner charges}
\label{sec:mappingformulae}
In this section, we provide explicit formulas for the corner charge in terms of the topological invariants as evaluated on the entire occupied subspace of a given model.
For systems with $\mathcal{I}$, $C_3$, and $C_3 + \mathcal{I}$ symmetry, we can uniquely identify the spinless limit of a given spinful model. In this case we can employ the results of Ref.~\onlinecite{benalcazar2018quantization}. In the remaining cases we deduce the formulas from the EBR tables given in section~\ref{sec:EBRdecomposition}. Importantly, all corner charges appearing in these formulas as well as in the EBR tables apply only to crystal terminations where $\Lambda_{\mathrm{F}}$ in Eq.~\eqref{eq:finitelattice} has corners at the intersection of 1D edges that are obtained from translating unit cells with crystal lattice vectors~\cite{benalcazar2018quantization}, but not necessarily primitive ones.

As noted before, in the case where we only have $C_4$ symmetry at our disposal, no corner charge formula can be constructed from our invariants. We leave the investigation of this symmetry class to future work, and instead consider the case of $C_4 + \mathcal{I}$ symmetry here.

\subsubsection{$\mathcal{I}$ symmetry}
Inversion symmetry becomes equal to $C_2$ symmetry in the spinless case. This means that, using inversion eigenvalues, we can uniquely read off the $C_2$ eigenvalues of the spinless version of any model at hand, and may then use the formula presented in Ref.~\onlinecite{benalcazar2018quantization} for spinless $C_2$ symmetry to infer the corner charge of our model. Note that the doubling of the corner charge, which comes with going from spinless to spinful and imposing TRS, is automatically taken into account by the fact that the inversion eigenvalues are equal for Kramers partners.
We therefore obtain
\begin{equation}
\label{eq:cornerchargewithjustinversion}
Q_\mathrm{c} = \frac{1}{4}\left([X^\mathcal{I}_2] + [Y^\mathcal{I}_2] - [M^\mathcal{I}_2]\right) \mod 2.
\end{equation}
A nonzero value implies two equal fractional corner charges at $\mathcal{I}$-related sectors with $Q_\mathrm{c} = 1$.

\subsubsection{$C_2$ symmetry}
Comparing with Table~\ref{tab:EBRc2}, we have
\begin{equation}
Q_\mathrm{c} = \nu_{x}^{\pi},
\end{equation}
where, if $H_{W_x} (k_y=0,\pi)$ does not have pinned bands at eigenvalue $\pi$, we declare $\nu_{x}^{\pi} = 0$. We note that $\nu_{x}^{\pi}$ is $\mathbb{Z}_2$ valued, in accordance with the fact that two Wannier Kramers pairs at $1b$ are trivial in that they can be removed from $1b$ and moved around the unit cell in a $C_2$ symmetric fashion. A nonzero value of $Q_\mathrm{c} = 1$ implies two equal fractional corner charges at $C_2$-related sectors.

\subsubsection{$C_3$ symmetry}
As shown in Appendix~\ref{sec:C3appendix}, there is a one-to-one mapping between the $C_3$ eigenvalues of the spinless and spinful cases. It implies that
\begin{equation}
Q_\mathrm{c} = \frac{2}{3} \left([K^{(3)}_1]+[K^{(3)}_2]\right) \mod 2.
\end{equation}
A nonzero value implies three equal fractional corner charges at $C_3$-related corners, with possibilities $Q_\mathrm{c} = \frac{2}{3}$ or $Q_\mathrm{c} = \frac{4}{3}$.

\subsubsection{$C_3 + \mathcal{I}$ symmetry}
The one-to-one mapping of $C_3$ eigenvalues from Appendix~\ref{sec:C3appendix}, as well the observation that inversion symmetry becomes the same as $C_2$ symmetry in the spinless case, yields
\begin{equation}
Q_\mathrm{c} = - \frac{1}{4} [M^\mathcal{I}_2] -\frac{1}{3} [K^{(3)}_2] \mod 2.
\label{eq:C3InvQc}
\end{equation}
A nonzero value implies six equal fractional corner charges at $C_3, \mathcal{I}$-related corners, with possibilities $Q_\mathrm{c} = \frac{1}{3}$, $Q_\mathrm{c} = \frac{2}{3}$, $Q_\mathrm{c} = 1$, $Q_\mathrm{c} = \frac{4}{3}$, or $Q_\mathrm{c} = \frac{5}{3}$.

\subsubsection{$C_4 + \mathcal{I}$ symmetry}

While $\mathcal{I}$ symmetry only allows for the decomposition of the sample into two halves, and therefore for a corner charge quantized in units of $1 \mod 2$, $C_4$ symmetry affords a further halving, so that the corner charge is quantized in units of $1/2 \mod 2$. Any $\mathcal{I}$ protected corner charge can in this way be split up into two $C_4 + \mathcal{I}$ protected corner charges of half the size. Using Eq.~\eqref{eq:cornerchargewithjustinversion}, we therefore obtain
\begin{equation}
\label{eq:cornerchargewithjustinversion}
Q_\mathrm{c} = \frac{[X^\mathcal{I}_2]}{4} - \frac{[M^\mathcal{I}_2]}{8} \mod 2,
\end{equation}
which we simplified by the $C_4$ constraint $[X^\mathcal{I}_2] = [Y^\mathcal{I}_2]$.
A nonzero value of $Q_\mathrm{c}$ implies four equal fractional corner charges at $C_4$-related corners (this configuration is automatically $\mathcal{I}$ symmetric), with possibilities $Q_\mathrm{c} = \frac{1}{2}$, $Q_\mathrm{c} = 1$, $Q_\mathrm{c} = \frac{3}{2}$.

\subsubsection{$C_6$ symmetry}
Comparing with Table~\ref{tab:EBRc6}, we have
\begin{equation}
Q_\mathrm{c} = \mu_{\bs{\Gamma M}} -\frac{1}{3} [K^{(3)}_2] \mod 2,
\end{equation}
where $\mu_{\bs{\Gamma M}}$ denotes the parity of the number of $W_{\bs{\Gamma M}}$ zero eigenvalue pairs. A nonzero value implies six equal fractional corner charges at $C_6$-related corners, with possibilities $Q_\mathrm{c} = \frac{1}{3}$, $Q_\mathrm{c} = \frac{2}{3}$, $Q_\mathrm{c} = 1$, $Q_\mathrm{c} = \frac{4}{3}$, or $Q_\mathrm{c} = \frac{5}{3}$.

\subsubsection{Summary}
\begin{table}[H]
\centering
\begin{tabular}{|c||c|}
\hline
$\mathcal{S}$ & $Q_\mathrm{c}$ \\ \hhline{|=#=|}
$\mathcal{I}$ & $\frac{1}{4}\left([X^\mathcal{I}_2] + [Y^\mathcal{I}_2] - [M^\mathcal{I}_2]\right)$ \\ \hline
$C_2$ & $\nu_{x}^{\pi}$ \\ \hline
$C_3$ & $\frac{2}{3} ([K^{(3)}_1]+[K^{(3)}_2])$ \\ \hline
$C_3 + \mathcal{I}$ & $- \frac{1}{4} [M^\mathcal{I}_2] -\frac{1}{3} [K^{(3)}_2]$ \\ \hline
$C_4 + \mathcal{I}$ & $\frac{[X^\mathcal{I}_2]}{4} - \frac{[M^\mathcal{I}_2]}{8}$ \\ \hline
$C_6$ & $\mu_{\bs{\Gamma M}} -\frac{1}{3} [K^{(3)}_2]$ \\ \hline
\end{tabular}
\caption{Summary of corner charge formulas.}
\end{table}

\section{Material candidates}
\label{sec:materials}
We propose the group-V buckled honeycomb monolayers of elemental antimony (Sb) and arsenic (As) as material realizations of protected fractional corner charges. Theoretical studies suggest that antimonene and arsenene can serve as an excellent platform for electronics due to high band gap tunability and mechanical stability~\cite{atomicVgroup, AsSbdft, AsTPTstrain, LattParamAs, AsSbthermoel}. Moreover, these 2D materials, as well as atomically thin bismuth monolayers (called bismuthene), deposited on a SiC substrate, are promising candidates for a realization of the quantum spin Hall states at room temperature~\cite{Murakami06,AsSbBiSiC1, AsSbBiSiC2}. Only recently, several experimental reports have demonstrated a successful fabrication of a monolayer structure of antimony~\cite{SbonAg, SbSyntMBA, SbonGe} and arsenic~\cite{AsAg111}.

\begin{figure}[H]
\centering
\includegraphics[width=0.8\textwidth]{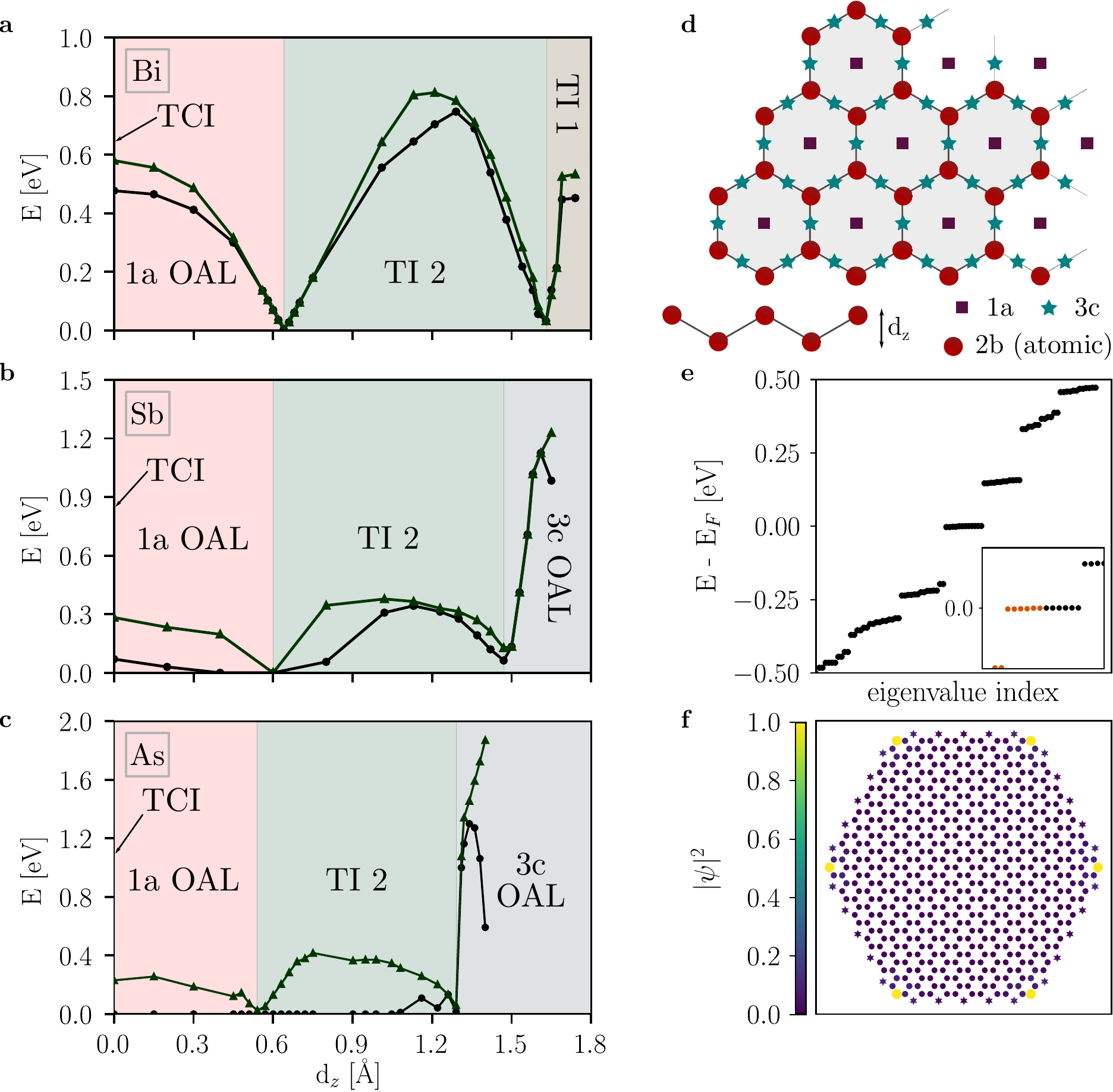}
\caption{Energy gap as a function of the buckling parameter $d_z$ for \textbf{a} bismuth, \textbf{b} antimony and \textbf{c} arsenic monolayers. The black line (circles) indicates the indirect gap, while the green line (triangles) indicates the direct gap. Top and side views of the lattice structure are illustrated in \textbf{d}, together with the Wyckoff positions of the space group 164. \textbf{e} Low-energy spectrum of a finite armchair-terminated flake of the $3c$ OAL. The inset presents the energies around the Fermi level, with filled states in orange. \textbf{f} The electronic densities of the corner states with color scale proportional to the normalized square modulus $|\psi_i|^2$ of the eigenstates (normalized with respect to the largest $|\psi_i|^2$). The tellurium atoms used for edge passivation are shown as stars.}
\label{fig:materials}
\end{figure}

Free-standing monolayers with nonzero buckling $d_z$ have a three-fold rotational symmetry $C_3$ as well as inversion $\mathcal{I}$ symmetry (consult Fig.~\ref{fig:materials}~\textbf{d}). (In practice, we consider weak substrate coupling so that the inversion symmetry is approximately retained.) Applying strain leads to a decreasing $d_z$ parameter up to a fully flat structure with six-fold symmetry. In Fig.~\ref{fig:materials}~\textbf{a}, \textbf{b}, \textbf{c}, we present the band gap evolution of Bi, Sb and As as a function of tensile strain, which is modeled by a modification to the in-plane lattice parameter (larger strain corresponds to a longer in-plane distance between atoms). First, we note the qualitative similarity of the phase diagrams for all three investigated materials. At $d_z$ = 0 (which corresponds to a large strain around $\sim$25\%), there is an additional mirror symmetry $M_z$, and all structures are in an topological crystalline insulating (TCI) phase, protected by a mirror Chern number, which we verified by Wilson loop calculations (not shown here). This phase does not have exponentially localized Wannier functions that respect all symmetries of the model. Small buckling breaks the mirror symmetry and the materials then realize an OAL with localized Wannier orbitals centered at the center of the hexagons in the honeycomb lattice (Wyckoff position $1a$ of the crystal). Upon further decreasing strain, a transition to a $\mathbb{Z}_2$ topological insulator (TI) is observed via a band gap closing around $d_z =0.6 \, \rm{\AA}$. To confirm this topological phase transition, we compute the $\mathbb{Z}_2$ topological index $\Delta_{\mathrm{TI}}$ given by the product of the inversion eigenvalues of the occupied bands at the time-reversal invariant momenta in the BZ~\cite{Fu07}, and obtain $\Delta_{\mathrm{TI}} = 1$. As strain decreases further, another band gap closing occurs. Here, the Bi monolayer reenters a TI phase (with different symmetry indicator invariants as shown in Table~\ref{tab:sym_indicators}), as confirmed by the $\mathbb{Z}_2$ index remaining nontrivial. In contrast, the almost fully buckled Sb and As monolayers enter once again in an OAL phase, this time with bands induced from the Wyckoff positions $3c$ (which is located on the bonds of the hexagon, see Fig.~\ref{fig:wyckpositions}~\textbf{d}). Hence, our results reveal more details on the previously investigated strain-induced topological phase transitions in these materials~\cite{BansilSbTriv,AsTPT, TPTVelem}.

Let us consider the systems with open boundary conditions. To establish the presence of corner charges, we perform open flake calculations for distinct OALs using the localized basis DFT method SIESTA~\cite{Soler_2002}. In Fig.~\ref{fig:materials}~\textbf{e}, \textbf{f}, we show results for a fully buckled antimony flake as a representative of the $3c$ OAL. The most direct indicator of fractional corner charges are corner-localized midgap states. If present, they are expected to appear close to the Fermi level. However, they are not necessarily well-separated from the bulk or edge modes. Therefore, we passivate the structure with tellurium atoms (marked with stars in Fig.~\ref{fig:materials}~\textbf{f}) in order to remove spurious dangling edge states from the bulk gap. The energy spectrum (see Fig.~\ref{fig:materials}~\textbf{e}) then exhibits 12 exactly degenerate corner states at the Fermi level, with only half of them filled. We thus obtain a fractional corner charge of $Q_\mathrm{c} = 1 \mod 2$ per corner, realizing a filling anomaly, as at the given filling it is not an insulating state that satisfies both charge neutrality and the crystalline symmetries.\cite{benalcazar2018quantization}

We confirm this corner charge using the topological indices developed in section~\ref{sec:topologicalindices}. In Table~\ref{tab:sym_indicators}, we evaluate the symmetry indicators for all discussed phases. We may then compute the corner charge of the $3c$ OAL on a hexagonal flake using Eq.~(\ref{eq:C3InvQc}). The relevant unit cell is the hexagonal cell, shown in Fig.~\ref{fig:wyckpositions}~\textbf{d}, which contains three primitive unit cells of the honeycomb lattice [space group 164 (P$\bar{3}$m1)]. The symmetry indicators in Table~\ref{tab:sym_indicators} are given for the primitive unit cell. To obtain the corresponding indicators for the hexagonal cell, we note that an enlargement of the unit cell results in a BZ folding, where the $K$ and $K'$ points are mapped onto $\Gamma$, while the $M$, $M'$ and $M''$ points are left unchanged. Referring to Table~\ref{tab:sym_indicators}, this implies $\chi^{(3)}_{\mathcal{I}} = (4,0)$ for the hexagonal cell, from which we obtain $Q_c = 1 \mod 2$ by Eq.~\eqref{eq:C3InvQc}. This is in agreement with the numerical results presented in Fig.~\ref{fig:materials}~\textbf{e}, \textbf{f}.

Correspondingly, in the case of the $1a$ OAL, we obtain $\chi^{(3)}_{\mathcal{I}} = (0,0)$ for the hexagonal cell (the primitive cell cannot be used to build a $C_3$-symmetric finite geometry). We conclude that there are no fractional charges. This is a case in point: although the $1a$ atomic limit is obstructed, in the sense that the electrons are localized away from the atomic sites, which are located at the $2b$ Wyckoff position of the crystal, there are no protected corner charges. (There may however be such charges in $C_3$-symmetric geometries that are terminated by cutting through unit cells. We do not consider these geometries here, mainly because there is no bulk-boundary correspondence in this case, and the actual corner charge is dependent on how the boundary unit cells are cut.)

\onecolumngrid

\twocolumngrid

\onecolumngrid

\begin{table}[b]
	\centering
	\addtolength\tabcolsep{5pt}
	\begin{tabular}{|c||c|c|c|c|c|c||c||c|}
		\hline 
		phase & $\# \Gamma_2^{\mathcal{I}}$&  $\# M_2^{\mathcal{I}}$ & $[M^\mathcal{I}_2]$ & $\# \Gamma_2^{(3)}$ & $\# K_2^{(3)} $ & $[K^{(3)}_2]$  &  $\chi^{(3)}_{\mathcal{I}} = ([M^\mathcal{I}_2], [K^{(3)}_2])$ & $\Delta_{\textnormal{TI}}$ \\
\hhline{|=#=|=|=|=|=|=#=#=|}
		TI 1 & 4 & 6 & 2 & 0 & 4 & 4  & (2, 4) & 1 \\
\hline 
		TI 2 & 4 & 6 & 2 & 2 & 4 & 2 & (2,  2) & 1 \\
		\hline 
		$3c$ OAL & 2 & 6 & 4 & 2 & 4 & 2  & (4, 2) & 0 \\
		\hline 
		$1a$ OAL & 4 & 4 & 0 & 2 & 4 & 2 & (0, 2) & 0\\
		\hline 
	\end{tabular} 
	\caption{Topological invariants and symmetry indicators $\chi^{(3)}_{\mathcal{I}}$ corresponding to different regions in the phase diagrams. The symmetry indicators were calculated using the primitive 2-site unit cell of the honeycomb lattice (see Table~\ref{tab:band_reps} for a decomposition in terms of elementary band representations). The indices $\chi^{(3)}_{\mathcal{I}}$ allow for a more refined classification even of the strong TIs. We find that the $3c$ and $1a$ OALs differ in their inversion indicator $[M^\mathcal{I}_2]$ and thus, as explained in the main text, inversion-symmetric flakes built from their hexagonal unit cells differ by a protected corner charge equal to $1 \mod 2$.}
	\label{tab:sym_indicators}
\end{table}

\newpage

\twocolumngrid

\section{Discussion}
As established in Refs.~\onlinecite{Miert16,miertOrtixFractionalCharge17,rhim17} for insulators with bulk polarization and more recently in Refs.~\onlinecite{miertcorners,benalcazar2018quantization} for second-order topological insulators, the nontrivial bulk topology of OALs can be revealed via charge fractionalization at boundaries. This represents the simplest mechanism for a topological bulk-boundary correspondence that is protected by crystalline symmetries. In this work, we presented theory and material candidates for charge fractionalization at corners in 2D systems with significant spin-orbit coupling, thus providing a broader picture than the one presented in some recent previous treatments of this phenomenon~\cite{miertcorners,EzawaWannier19,benalcazar2018quantization}. 

Corner charges in topological insulators  are well-defined when there is no edge spectral flow but also only in the absence of an edge-induced filling anomaly due to (time-reversal) bulk polarization. Since there is no crystalline symmetry-protected edge spectral flow in 2D (assuming the symmetry acts at least in part non-locally), corner charges are well defined for all 2D systems that are not strong or weak first-order topological insulators, or $M_z$ mirror Chern insulators~\cite{Fu11,Ando15}.

Diagnosing spinful OALs with time-reversal symmetry in 2D was particularly challenging because the irreducible representations of the occupied bands at HSPs are usually two-dimensional, yielding trivial symmetry indicator invariants at $C_2$-invariant HSPs. Symmetry indicators were therefore insufficient to identify the Wannier centers in $C_6$, $C_4$, and $C_2$ symmetric insulators. This is straightforwardly manifested in the fact that inducing a band representation from spinful maximal Wyckoff positions exhausts the symmetry representations at various HSPs of the BZ. To overcome this difficulty, we considered Wilson loop and nested Wilson loop invariants, which could better ``resolve'' the positions of the Wannier centers. Wilson loops, however, are essentially one dimensional objects that extract projections of the 2D positions of the Wannier centers along particular directions. Nested Wilson loops are a best-effort attempt to localize the Wannier centers in 2D, but cannot always be interpreted literally due to the possible non-commutation of Wilson loops along different directions. In the presence of crystalline symmetries, however, Wilson and nested Wilson loops have eigenvalues with quantized phases, which clearly distinguish different OALs in $C_6$ and $C_2$ symmetric insulators, but are insufficient for insulators which only have $C_4$ symmetry. We leave the challenge of finding a formula for the corner charge in such systems to future work.

We studied the protection due to only spatial symmetries because corner charge fractionalization is a robust observable that does not require additional spectral symmetries such as chiral or particle-hole symmetry. However, when particle-hole symmetry is present, we can  additionally predict topologically protected zero-energy corner states. These are characterized by $Q_\mathrm{c} = 1 \mod 2$: Consider a system with an $n$-fold symmetry in a phase with $2n$ degenerate midgap states (the 2 is due to TRS). At half-filling, $n$ midgap states are occupied and there is no gap. To arrive at a gapped system (as required for the corner charge to be well defined), we need to either fill $n$ more states or remove $n$ electrons from the charge-neutral system. When maintaining the crystal symmetry, this implies an excess (or missing) charge of $Q_\mathrm{c} = 1 \mod 2$ for each of the $n$ corners~\cite{benalcazar2018quantization}.

Interestingly, we find that there are obstructed atomic limits, where the electrons are localized away from the atomic sites, which still do not have nontrivial corner charges. These may instead be diagnosed by their response to crystal defects~\cite{benalcazar2018quantization,li2019fractional}. We leave the exploration of the defect response of obstructed atomic limits with significant spin-orbit coupling to future work.

\begin{acknowledgments}
We thank Aris Alexandradinata, Barry Bradlyn, Sander Kooi, Tianhe Li and Taylor Hughes for helpful discussions. F.~S., S.~S.~T., and T.~N. acknowledge support from the Swiss National Science Foundation (grant number: 200021\_169061) and from the European Union's Horizon 2020 research and innovation program (ERC-StG-Neupert-757867-PARATOP). F. S. was supported in part by the National Science Foundation under Grant No. NSF PHY-1748958 and by the Heising-Simons Foundation. M.~B. acknowledges financial support from the Polish NCN Grant No. 2014/14/A/ST3/00654 and by the statutory grant 0402/0016/18 of the Wroclaw University of Science and Technology. W.~A.~B. thanks the support of the Eberly Postdoctoral Fellowship at the Pennsylvania State University. M.~G.~V. acknowledges the IS2016-75862-P national project of the Spanish MINECO.
\end{acknowledgments}

\begin{widetext}
\appendix

\section{Consequences of rotation symmetry}
\label{sec:rotsymwladimir}
To obtain the constraints on the symmetry eigenvalues used in the main text, we here derive the consequences of rotational symmetry for the Bloch eigenstates of a crystal.
Rotation symmetry is expressed as
\begin{align}
\hat{r} h({\bs{k}}) \hat{r}^\dagger = h(R{\bs{k}}).
\label{eq:RotationSymmetry}
\end{align}
Here, $\hat{r}$ is the $n$-fold rotation operator (we could also write this operator as $\hat{r}_n$, but we will omit the subscript for simplicity) and $R$ is the matrix that rotates the crystal momentum by an angle of $\frac{2\pi}{n}$. For systems in class AII, the rotation operator obeys $\hat{r}^n=-1$. From \eqref{eq:RotationSymmetry}, it follows that
\begin{align}
h(R{\bs{k}}) \hat{r} \ket{u_{\bs{k}}^n}=\hat{r} h({\bs{k}})  \ket{u_{\bs{k}}^n}= \epsilon_n({\bs{k}}) \hat{r}\ket{u_{\bs{k}}^n}.
\end{align}
Thus, $\hat{r} \ket{u_{\bs{k}}^n}$ is an eigenstate of $h(R{\bs{k}})$ with energy $\epsilon_n({\bs{k}})$. This means that we can write the expansion
\begin{align}
\hat{r} \ket{u_{\bs{k}}^n} &= \sum_m \ket{u_{R{\bs{k}}}^m}\bra{u_{R{\bs{k}}}^m} \hat{r} \ket{u_{\bs{k}}^n}\nonumber\\
&=\sum_m \ket{u_{R{\bs{k}}}} B_{\bs{k}}^{mn},
\end{align}
where
\begin{align}
B^{mn}_{\bs{k}}&=\bra{u_{R{\bs{k}}}^m} \hat{r} \ket{u_{\bs{k}}^n}
\end{align}
is the sewing matrix, which is unitary:
\begin{align}
B^{ml}_{\bs{k}} (B^{\dagger}_{\bs{k}})^{ln}&=\bra{u_{R{\bs{k}}}^m} \hat{r} \ket{u_{\bs{k}}^{l}}\bra{u_{{\bs{k}}}^{l}} \hat{r}^\dagger \ket{u_{R{\bs{k}}}^n}\nonumber\\
&=\bra{u_{R{\bs{k}}}^m}\hat{r} \hat{r}^\dagger \ket{u_{R{\bs{k}}}^n}\nonumber\\
&=\delta_{mn}.
\end{align}
As before, let us use \eqref{eq:RotationSymmetry} to do the following calculation:
\begin{align}
h(R{\bs{k}})\hat{r} \ket{u_{\bs{k}}^n}&=\epsilon_n({\bs{k}}) \hat{r} \ket{u_{\bs{k}}^n}=\epsilon_n({\bs{k}}) \sum_m \ket{u_{R{\bs{k}}}^m} B_{\bs{k}}^{mn} \nonumber\\
&=h(R{\bs{k}})\sum_m \ket{u_{R{\bs{k}}}^m} B_{\bs{k}}^{mn}\nonumber\\
&=\sum_m \epsilon_m(R{\bs{k}}) \ket{u_{R{\bs{k}}}^m} B_{\bs{k}}^{mn},
\end{align}
from which it follows that
\begin{align}
\sum_m \ket{u_{R{\bs{k}}}^m} B_{\bs{k}}^{mn} (\epsilon_n({\bs{k}})-\epsilon_m(R{\bs{k}}))=0.
\end{align}
for every $n$. Since the eigenstates form an orthonormal basis, the expression above implies that
\begin{align}
B_{\bs{k}}^{mn} (\epsilon_n({\bs{k}})-\epsilon_m(R{\bs{k}}))=0.
\label{eq:aux1}
\end{align}
for every $m$ and $n$. Equation~\eqref{eq:aux1} implies that the sewing matrix $B_{\bs{k}}^{mn}$ only connects states at ${\bs{k}}$ and $R{\bs{k}}$ having the same energy. 

\section{Invariant points under rotation}
\label{sec:invptswladimir}

Now we focus on the high symmetry points of the BZ (HSPs). These are points that obey
\begin{align}
R {\bf \Pi} = {\bf \Pi}
\end{align}
up to a reciprocal lattice vector. 
These points are shown in Fig.~\ref{fig:BZ} for all the $C_{n=2,3,4,6}$ symmetries. At HSPs, Eq.~\eqref{eq:RotationSymmetry} reduces to $\hat{r} h({\bf \Pi}) \hat{r}^\dagger = h({\bf \Pi})$, where $\hat{r}$ here corresponds to the rotation operator of the little group at the HSP ${\bf \Pi}$. This expression is compactly written as
\begin{align}
[\hat{r},h({\bf \Pi})]=0.
\end{align}
Thus, it is possible to choose a gauge in which the energy eigenstates are also eigenstates of the rotation operator,
\begin{align}
\hat{r} \ket{u_{\bf \Pi}^n}=r_{\bf \Pi}^n \ket{u_{\bf \Pi}^n}.
\end{align}
This is automatic if there are no degeneracies, but if energy degeneracies exist, one can always choose a gauge such that the above expression is possible. At these invariant points, the sewing matrix is diagonal:
\begin{align}
B_{\bf \Pi}^{mn}=\bra{u_{\bf \Pi}^m} \hat{r} \ket{u_{\bf \Pi}^n}=r^n_{\bf \Pi} \braket{u_{\bf \Pi}^m}{u_{\bf \Pi}^n}=r^n_{\bf \Pi} \delta_{mn}.
\end{align}

Now, we show that the rotation eigenvalues of HSPs that are related by symmetry are equal. Consider the rotation by an angle $\phi$ in a crystal with $C_{2\pi/\phi}$ symmetry. This rotation symmetry relates HSPs that are invariant under rotations by a larger angle $\theta=n \phi$, for $n$ integer. Call these HSPs ${\bf \Pi}_\theta$. Here, we are interested in knowing how the rotation eigenvalues of ${\bf \Pi}_\theta$ and $R_\phi {\bf \Pi}_\theta$ are related. In particular, this applies to two cases: (i) In $C_6$-symmetric crystals, $\phi=2\pi/6$. For $\theta_1=2\pi/3=2\phi$ we have $\bs{K}=R_\phi \bs{K'}$, while for $\theta_2=\pi=3\phi$ we have $\bs{M'}=R_\phi \bs{M}=R_\phi^2 \bs{M''}$; (ii) in $C_4$-symmetric crystals, $\phi=\pi/2$, for $\theta=\pi=2\phi$ we have $\bs{X'}=R_\phi \bs{X}$.
Let us start by asking what we get from applying $\hat{r}_\theta \ket{u_{{R_\phi}{\bf \Pi_\theta}}^n}$. Since ${R_\phi}{\bf \Pi_\theta}$ is invariant under $\hat{r}_\theta$, we have
\begin{align}
\hat{r}_\theta \ket{u_{{R_\phi}{\bf \Pi_\theta}}^n}=r_{R_\theta {\bf \Pi_\theta}}^n \ket{u_{{R_\phi}{\bf \Pi_\theta}}^n}.
\label{eq:app_rotation_eigenvalues}
\end{align}
Since ${R_\phi}{\bf \Pi_\theta}$ and ${\bf \Pi_\theta}$ are related by $C_{2\pi/\phi}$ symmetry, we can expand
\begin{align}
\hat{r}_\phi \ket{u_{{\bf \Pi}_\theta}^n}&=\sum_m \ket{u_{R_\phi {\bf \Pi}_\theta}^m} \bra{u_{R_\phi {\bf \Pi}_\theta}^m}\hat{r}_\phi \ket{u_{{\bf \Pi}_\theta}^n}\nonumber\\
&=\sum_m \ket{u_{R_\phi {\bf \Pi}_\theta}^m} B_{{\bf \Pi}_\theta}^{mn},
\label{eq:app_expansion}
\end{align}
where $B_{{\bf \Pi}_\theta}^{mn}=\bra{u_{R_\phi {\bf \Pi}_\theta}^m}\hat{r}_\phi \ket{u_{{\bf \Pi}_\theta}^n}$ is the sewing matrix, with the properties shown before. Conversely, we also have that
\begin{align}
\ket{u_{R_\phi {\bf \Pi}_\theta}^n}=\sum_m\hat{r}_\phi \ket{u_{{\bf \Pi}_\theta}^m}\left[B_{{\bf \Pi}_\theta}^{\dagger}\right]^{mn}.
\label{eq:app_expansion_inverse}
\end{align}
So, replacing this expansion in \eqref{eq:app_rotation_eigenvalues}, we have
\begin{align}
\hat{r}_\theta \ket{u_{R_\phi {\bf \Pi}_\theta}^n}= \hat{r}_\phi \sum_m r_{R_\theta {\bf \Pi_\theta}}^n \ket{u_{{\bf \Pi}_\theta}^m}\left[B_{{\bf \Pi}_\theta}^{\dagger}\right]^{mn}.
\label{eq:app_rotation_eigenvalues_exp1}
\end{align}
Taking a different approach, we calculate directly the rotation eigenvalues in the expansion \eqref{eq:app_expansion_inverse} to get
\begin{align}
\hat{r}_\theta \ket{u_{R_\phi {\bf \Pi}_\theta}^n} &= \hat{r}_\theta \sum_m\hat{r}_\phi \ket{u_{{\bf \Pi}_\theta}^m}\left[B_{{\bf \Pi}_\theta}^{\dagger}\right]^{mn}\nonumber\\
&= \hat{r}_\phi \sum_m \hat{r}_\theta \ket{u_{{\bf \Pi}_\theta}^m}\left[B_{{\bf \Pi}_\theta}^{\dagger}\right]^{mn}\nonumber\\
&= \hat{r}_\phi \sum_m r_{{\bf \Pi}_\theta}^m \ket{u_{{\bf \Pi}_\theta}^m}\left[B_{{\bf \Pi}_\theta}^{\dagger}\right]^{mn}.
\end{align}
So, comparing the last two results we conclude that
\begin{align}
\sum_m (r_{R_\phi {\bf \Pi}_\theta}^n-r_{{\bf \Pi}_\theta}^m) \ket{u_{{\bf \Pi}_\theta}^m}\left[B_{{\bf \Pi}_\theta}^{\dagger}\right]^{mn}=0.
\end{align}
for all $n$. Furthermore, since the eigenstates form an orthonormal basis, we have
\begin{align}
(r_{R_\phi {\bf \Pi}_\theta}^n-r_{{\bf \Pi}_\theta}^m) \left[B_{{\bf \Pi}_\theta}^{\dagger}\right]^{mn}=0,
\end{align}
for all $m$ and $n$. Now, the sewing matrix will have non-zero elements for equal energies at the two different HSPs $R_\phi {\bf \Pi}_\theta$ and ${\bf \Pi}_\theta$. Thus, for $\epsilon_m(R_\phi {\bf \Pi}_\theta)=\epsilon_n({\bf \Pi}_\theta)$, we need $r^m_{{\bf \Pi}_\theta}=r^n_{R_\phi {\bf \Pi}_\theta}$, i.e., the rotation spectra at $R_\phi {\bf \Pi}_\theta$ and ${\bf \Pi}_\theta$ are equal for bands having equal energies.
In particular, we have the relations
\begin{align}
\{r^n_{\bs{K}}\}&\stackrel{C_6}{=}\{r^n_{\bs{K'}}\},\nonumber\\
\{r^n_{\bs{M}}\}& \stackrel{C_6}{=} \{r^n_{\bs{M'}}\} \stackrel{C_6}{=} \{r^n_{\bs{M''}}\},\nonumber\\
\{r^n_{\bs{X}}\}& \stackrel{C_4}{=} \{r^n_{\bs{X'}}\}.
\end{align}

\section{Constraints on the rotation eigenvalues due to time-reversal symmetry}
\label{sec:trscnstrnswladimir}

Now we look at the interplay between TRS and rotation symmetry. The two operators commute:
\begin{align}
[\Theta, \hat{r}]=0.
\end{align}
Thus, on one hand we have
\begin{align}
\Theta \left( \hat{r} \ket{u_{{\bs{k}}}^l} \right) &= \Theta \left( \sum_n \ket{u_{R {\bs{k}}}^n} B_{\bs{k}}^{nl} \right)\nonumber\\
&= \sum_{m,n}\ket{u_{-R {\bs{k}}}^m} V_{R {\bs{k}}}^{mn} B_{\bs{k}}^{nl*},
\end{align}
where $V$ is the sewing matrix for TRS. On the other hand, we have
\begin{align}
\hat{r} \left( \Theta \ket{u_{{\bs{k}}}^l} \right) &= \hat{r} \left( \sum_m \ket{u_{- {\bs{k}}}^n} V_{\bs{k}}^{nl} \right)\nonumber\\
&= \sum_{m,n} \ket{u_{-R {\bs{k}}}^m} B_{- {\bs{k}}}^{mn} V_{\bs{k}}^{nl}.
\end{align}
In the last expression, we have used the fact that $R (- {\bs{k}})= - R {\bs{k}}$. From these two expressions we conclude that
\begin{align}
\sum_{m,n} \ket{u_{-R {\bs{k}}}^m} \left( V_{R {\bs{k}}}^{mn} B_{\bs{k}}^{nl*}  - B_{- {\bs{k}}}^{mn} V_{\bs{k}}^{nl} \right) = 0
\end{align}
for all $l$. Since the eigenstates are orthonormal, this relation implies that
\begin{align}
\sum_{n} \left( V_{R {\bs{k}}}^{mn} B_{\bs{k}}^{nl*}  - B_{- {\bs{k}}}^{mn} V_{\bs{k}}^{nl} \right) = 0
\end{align}
for all $m$, $l$. As noted earlier, of particular interest are the HSPs. At these points, $B_{{\bf \Pi}}^{mn}=r^n_{{\bf \Pi}}\delta_{mn}$ in the gauge in which $\{ \ket{u^n_{{\bf \Pi}}}\}$ are rotation eigenstates. Then, at these points, the previous relation results in
\begin{align}
V_{{\bf \Pi}}^{ml} \left( r^{l*}_{{\bf \Pi}} - r^m_{-{\bf \Pi}} \right) = 0
\end{align}
for all $l$, $m$. Thus, if $V_{{\bf \Pi}}^{ml} \neq 0$, $r^{l*}_{{\bf \Pi}} = r^m_{-{\bf \Pi}}$. This is possible only if $\epsilon_m(-{\bf \Pi})=\epsilon_l({\bf \Pi})$. Thus, we have that, under TRS,
\begin{align}
\{r^n_{\bf \Pi}\}& \stackrel{\mathrm{TRS}}{=} \{r^{n*}_{-{\bf \Pi}}\}.
\end{align}
More specifically, for equal energies at ${\bs{k}}={\bf \Pi}$ and ${\bs{k}}=-{\bf \Pi}$, their rotation eigenvalues are complex conjugates of each other [if, on the other hand, $\epsilon_m(-{\bf \Pi}) \neq \epsilon_l({\bf \Pi})$, we have that $V_{{\bf \Pi}}^{ml}=0$, which means that there is no restriction on the rotation eigenvalues]. In particular, at TRIMs which are also HSPs, ${\bf \Pi}=-{\bf \Pi}$, we have that $r^{l*}_{{\bf \Pi}}=r^m_{{\bf \Pi}}$ for equal energies $\epsilon_m({\bf \Pi})= \epsilon_l({\bf \Pi})$. This imposes the following constraints on the rotation eigenvalues: (i) for a non-degenerate state labeled by $n$, $r^{n*}_{{\bf \Pi}}=r^{n}_{{\bf \Pi}}$, i.e., its rotation eigenvalue is real: $r^{n}_{{\bf \Pi}}=\pm 1$ and (ii) for two degenerate states $n=1,2$ one could have $r^1_{{\bf \Pi}}=\lambda$ and $r^2_{{\bf \Pi}}=\lambda^*$, so that $r^{1*}_{{\bf \Pi}}=\lambda^*=r^{2}_{{\bf \Pi}}$ and $r^{2*}_{{\bf \Pi}}=\lambda=r^{1}_{{\bf \Pi}}$, that is, in energy-degenerate states, the rotation eigenvalues can be complex, but have to come in complex conjugate pairs. As said before, these constraints follow for HSPs that are also TRIM. This is the case for all the HSPs except  $\bs{K}$ and $\bs{K'}$, which map into each other under TRS. 

\section{Mapping between spinless and spinful $C_3$ eigenvalues}
\label{sec:C3appendix}
We start with the spinless indicators
\begin{align}
[\tilde{K}^{(3)}_i]= \#\tilde{K}^{(3)}_i - \#\tilde{\Gamma}^{(3)}_i,
\end{align}
where $\tilde{K}^{(3)}_{i=1,2,3}$, $\tilde{\Gamma}^{(3)}_{i=1,2,3}=\{1,e^{i 2\pi/3},e^{-i 2\pi/3}\}$. Upon introducing spin, each spinless eigenvalue $\lambda$ contributes two spinful eigenvalues $\lambda e^{\pm \mathrm{i} \pi/3}$. From this we obtain the relations
\begin{equation}
\begin{aligned}
&[K^{(3)}_1]=[\tilde{K}^{(3)}_1]+[\tilde{K}^{(3)}_2], \\
&[K^{(3)}_2]=[\tilde{K}^{(3)}_2]+[\tilde{K}^{(3)}_3], \\
&[K^{(3)}_3]=[\tilde{K}^{(3)}_3]+[\tilde{K}^{(3)}_1],
\end{aligned}
\end{equation}
where the $[K^{(3)}_i]$, $i=1,2,3$, are defined in Eq.~\eqref{eq:C3spinfulindicators}. Together with the constraints in Eq.~\eqref{eq:secondsetofC3indicatorconstraints} this implies
\begin{equation}
\begin{aligned}
&[\tilde{K}^{(3)}_1]=-[K^{(3)}_2], \\
&[\tilde{K}^{(3)}_2]=-[K^{(3)}_3], \\
&[\tilde{K}^{(3)}_3]=-[K^{(3)}_1],
\end{aligned}
\end{equation}
providing a mapping between spinless and spinful $C_3$ eigenvalues.

\section{Induction of band representations from maximal Wyckoff positions and relation to symmetry indicator invariants}
\label{sec:inductionEBRs}
In this Section, we explicitly induce the energy band representations at HSPs of the BZ following the prescription in Ref.~\onlinecite{Cano17-2}. Given a site symmetry representation, the induced band representation will allow us to identify the symmetry indicator invariants associated with a maximal Wyckoff position. In this section, we induce the band representations and corresponding symmetry indicator invariants for all the allowed site symmetry representations of spinful time-reversal symmetric orbitals at each maximal Wyckoff position.
In the following, $\rho$ refers to the representation of the site symmetry group, while $\rho^{\bs{k}}_G$ refers to the band representation at crystal momentum $\bs{k}$. We treat each case separately.

For $C_4$ and $C_2$ symmetries, we use the following primitive vectors $a_1=(1,0)$, $a_2=(0,1)$, and for both $C_6$ and $C_3$ symmetries, we use the following primitive vectors $a_1=(1,0)$, $a_{2,3}=(\pm \frac{1}{2}, \frac{\sqrt{3}}{2})$.

\subsection{$C_4$ symmetry: Representations induced from $2c$}
Given a site symmetry representation $\rho(C_2)$ of the orbitals at $2c$, the band representations are
\begin{align}
\rho_G^{\bs{k}}(C_4)&=\left(\begin{array}{cc}
0 & e^{\mathrm{i} {\bs{k}}.{a}_1}\rho(C_2)\\
1 & 0
\end{array}\right), \\
\rho_G^{\bs{k}}(C_2)&=\left(\begin{array}{cc}
e^{\mathrm{i} {\bs{k}}.{a}_1} & 0\\
0 & e^{\mathrm{i} {\bs{k}}.{a}_2}
\end{array}\right) \rho(C_2).
\end{align}

Let us consider one the only possible site symmetry representation, $\rho(C_2)=e^{\mathrm{i} \frac{\pi}{2} \sigma_z}$. 
For $C_4$, the band representations at HSPs are
\begin{align}
\rho_G^{\bs{\Gamma}}(C_4)=\left(\begin{array}{cc}
0 & e^{\mathrm{i} \frac{\pi}{2} \sigma_z}\\
1 & 0\end{array}\right), \quad
\rho_G^{\bs{M}}(C_4)=\left(\begin{array}{cc}
0 & -e^{\mathrm{i} \frac{\pi}{2} \sigma_z}\\
1 & 0\end{array}\right).\nonumber
\end{align}
Both of these matrices have the four eigenvalues $e^{\mathrm{i} \pi/4}$, $e^{-\mathrm{i} \pi/4}$, $e^{3\mathrm{i} \pi/4}$, $e^{-3\mathrm{i} \pi/4}$. Therefore, $[M_1^{(4)}]=0$.

For $C_2$, the band representations at HSPs are
\begin{align}
\rho_G^{\bs{\Gamma}}(C_2)=\sigma_0 e^{\mathrm{i} \frac{\pi}{2} \sigma_z}, \quad
\rho_G^{\bs{X}}(C_2)=-\sigma_z e^{\mathrm{i} \frac{\pi}{2} \sigma_z},\nonumber \\
\rho_G^{\bs{Y}}(C_2)=\sigma_z e^{\mathrm{i} \frac{\pi}{2} \sigma_z}, \quad
\rho_G^{\bs{M}}(C_2)=-\sigma_0 e^{\mathrm{i} \frac{\pi}{2} \sigma_z}. \nonumber
\end{align}
All these matrices have eigenvalues $+i$,$+i$, $-i$, $-i$, also leading to vanishing symmetry indicators. As we will see, this is also the case when the band representations are induced from $1b$: in fact, with spinful time-reversal symmetry, the only possible EBR is given by a pair of states with $C_2$ eigenvalues ($+i$, $-i$). Therefore, no symmetry indicators exist associated with the band representations of $C_2$.

\subsection{$C_4$ symmetry: Representations induced from $1b$}
Given a site symmetry representation $\rho(C_4)$ of the orbitals at $1b$, the band representations are
\begin{align}
\rho_G^{\bs{k}}(C_4)&=
e^{\mathrm{i} {\bs{k}}.{a}_1}\rho(C_4),\\
\rho_G^{\bs{k}}(C_2)&=e^{\mathrm{i} {\bs{k}}.(a_1+a_2)}\rho(C_2).
\end{align}
where $\rho(C_2)=\rho^2(C_4)$.
Let us consider the site symmetry representation $\rho(C_4)=e^{\mathrm{i} \frac{\pi}{4} \sigma_z}$.
For $C_4$, the band representations at HSPs are
\begin{align}
\rho_G^{\bs{\Gamma}}(C_4)= e^{\mathrm{i} \frac{\pi}{4} \sigma_z}, \quad
\rho_G^{\bs{M}}(C_4)=- e^{\mathrm{i} \frac{\pi}{4} \sigma_z}.\nonumber 
\end{align}
The matrix for the band representation of $C_4$ at ${\bs{\Gamma}}$ has eigenvalues $e^{\mathrm{i} \pi/4}$, $e^{-\mathrm{i} \pi/4}$, while the one at ${\bs{M}}$ has eigenvalues $e^{3\mathrm{i} \pi/4}$, $e^{-3\mathrm{i} \pi/4}$. Thus, $[M_1^{(4)}]=1$. Now, if the site symmetry representation were $\rho(C_4)=-e^{\mathrm{i} \frac{\pi}{4} \sigma_z}$ instead, the band representations at ${\bs{\Gamma}}$ and ${\bs{M}}$ would flip. This leads to the symmetry indicator invariant $[M_1^{(4)}]=-1$.

For $C_2$, the band representations are always of the form $\pm e^{\mathrm{i} \frac{\pi}{2} \sigma_z}$, which has eigenvalues $+i$, $-i$, leading to vanishing symmetry indicators. 

Let us now consider obstructions arising from the band representation when multiple orbitals localize at $1b$. If the two orbitals have the same representation, e.g., $\rho(C_4)=e^{\mathrm{i} \frac{\pi}{4} \sigma_z}$, the overall site symmetry representation, $\sigma_0 e^{\mathrm{i} \frac{\pi}{4} \sigma_z}$, induces a band representation with invariant $[M_1^{(4)}]=2$. If, on the other hand, the representations at the two orbitals differ, the induced band representations will have an invariant $[M_1^{(4)}]=0$. Both cases, however, are obstructed, because it is not possible to smoothly move two Kramers pairs from $1b$ to $1a$ in a $C_4$ symmetric way. We see from this analysis that other invariants must exist beyond symmetry indicators that capture the obstruction of the case of two Kramers pairs with $[M_1^{(4)}]=0$. 

\subsection{$C_6$ symmetry: Representations induced from $2b$}
Given a site symmetry representation $\rho(C_3)$ of the orbitals at $2b$, the band representations are
\begin{align}
\rho_G^{\bs{k}}(C_3)&=\left(\begin{array}{cc}
e^{\mathrm{i} {\bs{k}}.{a}_1} & 0\\
0 & e^{-\mathrm{i} {\bs{k}}.{a}_1}
\end{array}\right) \rho(C_3),\\
\rho_G^{\bs{k}}(C_2)&=\left(\begin{array}{cc}
0 & -1\\
1 & 0
\end{array}\right) \mathbb{1}_{2N \times 2N},
\end{align}
where $N$ is the number of Kramers pairs in the site $2b$.
Consider one Kramers pair at $2b$.
For $C_3$, the band representations at the HSPs are
\begin{align}
\rho_G^{\bs{\Gamma}}(C_3)=\rho(C_3), \quad
\rho_G^{\bs{K}}(C_3)=e^{\mathrm{i}\frac{2\pi}{3} \sigma_z}\rho(C_3).\nonumber
\end{align}
Thus, for the site symmetry representation $\rho(C_3)=e^{\mathrm{i} \frac{n\pi}{3} \sigma_z}$ (for $n$=1or 3), the eigenvalues are $e^{\mathrm{i} \frac{n\pi}{3}}$, $e^{-\mathrm{i} \frac{n\pi}{3}}$ at ${\bs{\Gamma}}$, and $e^{\mathrm{i} \frac{\pi}{3} (n+2)}$, $e^{-\mathrm{i} \frac{\pi}{3} (n+2)}$ at ${\bs{K}}$. This yields the invariants in Table~\ref{tab:C6_inducedC3InvariantsFrom2b}.

\begin{table}
	\centering
	\begin{tabular}{|c|c|c|l|}
		\hline 
		Site symm. & evals ${\bs{\Gamma}}$ & evals ${\bs{K}}$ & Invariants\\
		\hline\hline
		& $\#\Gamma_1=2$ & $\#K_1=1$ & $[K_1^{(3)}]=-1$ \\
		$e^{\mathrm{i}\frac{\pi}{3}\sigma_z}$ & $\#\Gamma_2=0$ & $\#K_2=2$ & $[K_2^{(3)}]=2$ \\
		 & $\#\Gamma_3=2$ & $\#K_3=1$ & $[K_3^{(3)}]=-1$\\
		\hline
		& $\#\Gamma_1=0$ & $\#K_1=2$ & $[K_1^{(3)}]=2$ \\
		$-\sigma_0$ & $\#\Gamma_2=4$ & $\#K_2=0$ & $[K_2^{(3)}]=-4$ \\
		 & $\#\Gamma_3=0$ & $\#K_3=2$ & $[K_3^{(3)}]=2$\\
		\hline
	\end{tabular} 
	\caption{$C_6$ symmetry: $C_3$ invariants induced from Wyckoff position $2b$ with different site symmetry representations.}
	\label{tab:C6_inducedC3InvariantsFrom2b}
\end{table}

Notice, from the invariants in Table~\ref{tab:C6_inducedC3InvariantsFrom2b}, that the obstruction is lifted only if three Kramers pairs locate at $2b$, two with representations $e^{\mathrm{i}\frac{\pi}{3}\sigma_z}$ and one with $-\sigma_0$. This illustrates the fact that the number of Kramers pairs at a maximal Wyckoff position alone does not determine whether an OAL is trivial. The site symmetry representation is crucial; they determine whether the Kramers pairs are free to move symmetrically or not. 

Regarding $C_2$, it follows from the lack of dependence of $\rho_G^{\bs{k}}(C_2)$ on the crystal momentum, that all invariants vanish.

\subsection{$C_6$ symmetry: Representations induced from $3c$}
Given a site symmetry representation $\rho(C_2)$ of the orbitals at $3c$, the band representations are
\begin{align}
\rho_G^{\bs{k}}(C_3)&=\left(\begin{array}{ccc}
0 & 0 & -1\\
1 & 0 & 0\\
0 & 1 & 0
\end{array}\right) \mathbb{1}_{2N \times 2N},\\
\rho_G^{\bs{k}}(C_2)&=\left(\begin{array}{ccc}
e^{\mathrm{i} {\bs{k}}.{a}_2}& 0\\
0 & e^{-\mathrm{i} {\bs{k}}.{a}_1} &0\\
0 & 0 & e^{-\mathrm{i} {\bs{k}}.{a}_3}
\end{array}\right) \rho(C_2),
\end{align}
where $N$ is the number of Kramers pairs in the site $3c$.
For C3, the band representation is constant across the $C_3$ invariant HSPs. Therefore, all invariants are trivial. 
For $C_2$, the band representations at the HSPs are
\begin{align}
\rho_G^{\bs{\Gamma}}(C_2)=\mathbb{1}_{2 \times 2}\rho(C_2), \quad \rho_G^{\bs{M}}(C_2)=\left(\begin{array}{ccc}
-1 & 0 & 0\\
0 & -1 & 0\\
0 & 0 & 1
\end{array}\right)\rho(C_2). \nonumber
\end{align}
But since the site representation for a Kramers pair, $\rho(C_2)=e^{\mathrm{i} \frac{\pi}{2}\sigma_z}$, has eigenvalues $+i$, $-i$, the band representations at $\bs{\Gamma}$ and ${\bs{M}}$ are the same. 

\subsection{$C_3$ symmetry: Representations induced from $1b$}
Given a site symmetry representation $\rho(C_3)$ of the orbitals at $1b$, the band representations are
\begin{align}
\rho_G^{\bs{k}}(C_3)= e^{\mathrm{i} {\bs{k}}.a_2} \rho(C_3).
\end{align}

The band representations at the HSPs are

\begin{align}
\rho_G^{\bs{\Gamma}}(C_3)=\rho(C_3), \quad \rho_G^{\bs{K}}(C_3)=e^{\mathrm{i} \frac{2\pi}{3}}\rho(C_3), \quad \rho_G^{\bs{K}'}(C_3)=e^{-\mathrm{i} \frac{2\pi}{3}}\rho(C_3).\nonumber
\end{align}
Thus, the invariants depend on the site symmetry representation $\rho(C_3)$. They are shown in Table~\ref{tab:C3_inducedC3InvariantsFrom1b}. Since TRS relates ${\bs{K}}$ with ${\bs{K}'}$, we only provide the representations at $\bs{\Gamma}$ and ${\bs{K}}$.

\begin{table}
	\centering
	\begin{tabular}{|c|c|c|l|}
		\hline 
		Site symm. & evals ${\bs{\Gamma}}$ & evals ${\bs{K}}$ & Invariants\\
		\hline\hline
		& $\#\Gamma_1=1$ & $\#K_1=1$ & $[K_1^{(3)}]=0$ \\
		$e^{\mathrm{i}\frac{\pi}{3}\sigma_z}$ & $\#\Gamma_2=0$ & $\#K_2=1$ & $[K_2^{(3)}]=1$ \\
		 & $\#\Gamma_3=1$ & $\#K_3=0$ & $[K_3^{(3)}]=-1$\\
		\hline
		& $\#\Gamma_1=0$ & $\#K_1=0$ & $[K_1^{(3)}]=0$ \\
		$-\sigma_0$ & $\#\Gamma_2=2$ & $\#K_2=0$ & $[K_2^{(3)}]=-2$ \\
		 & $\#\Gamma_3=0$ & $\#K_3=2$ & $[K_3^{(3)}]=2$\\
		\hline
	\end{tabular} 
	\caption{$C_3$ symmetry: $C_3$ invariants induced from Wyckoff position $1b$ with different site symmetry representations.}
	\label{tab:C3_inducedC3InvariantsFrom1b}
\end{table}

Note that, in order to have a trivial insulator with three movable Kramers pairs at $1b$, two of them need to have the representation $e^{\mathrm{i}\frac{\pi}{3}\sigma_z}$ and the third one the representation $-\sigma_0$.

\subsection{$C_3$ symmetry: Representations induced from $1c$}
Given a site symmetry representation $\rho(C_3)$ of the orbitals at $1c$, the band representations are
\begin{align}
\rho_G^{\bs{k}}(C_3)= e^{\mathrm{i} {\bs{k}}.a_1} \rho(C_3).
\end{align}

The band representations at the HSPs are

\begin{align}
\rho_G^{\bs{\Gamma}}(C_3)=\rho(C_3), \quad \rho_G^{\bs{K}}(C_3)=e^{-\mathrm{i} \frac{2\pi}{3}}\rho(C_3), \quad \rho_G^{\bs{K}'}(C_3)=e^{\mathrm{i} \frac{2\pi}{3}}\rho(C_3).\nonumber
\end{align}
So, just as for $1b$, the invariants depend on the site symmetry representation $\rho(C_3)$. They are shown in Table~\ref{tab:C3_inducedC3InvariantsFrom1c}.

\begin{table}
	\centering
	\begin{tabular}{|c|c|c|l|}
		\hline 
		Site symm. & evals ${\bs{\Gamma}}$ & evals ${\bs{K}}$ & Invariants\\
		\hline\hline
		& $\#\Gamma_1=1$ & $\#K_1=0$ & $[K_1^{(3)}]=-1$ \\
		$e^{\mathrm{i}\frac{\pi}{3}\sigma_z}$ & $\#\Gamma_2=0$ & $\#K_2=1$ & $[K_2^{(3)}]=1$ \\
		 & $\#\Gamma_3=1$ & $\#K_3=1$ & $[K_3^{(3)}]=0$\\
		\hline
		& $\#\Gamma_1=0$ & $\#K_1=2$ & $[K_1^{(3)}]=2$ \\
		$-\sigma_0$ & $\#\Gamma_2=2$ & $\#K_2=0$ & $[K_2^{(3)}]=-2$ \\
		 & $\#\Gamma_3=0$ & $\#K_3=0$ & $[K_3^{(3)}]=0$\\
		\hline
	\end{tabular} 
	\caption{$C_3$ symmetry: $C_3$ invariants induced from Wyckoff position $1c$ with different site symmetry representations.}
	\label{tab:C3_inducedC3InvariantsFrom1c}
\end{table}

Just as before, in order to have a trivial insulator with three movable Kramers pairs at $1c$, two of them need to have the representation $e^{\mathrm{i}\frac{\pi}{3}\sigma_z}$ and the third one the representation $-\sigma_0$.

\subsection{$C_2$ symmetry: Representations induced from any $C_2$ invariant HSP}
Since all $C_2$ invariants points are also TRIM points, and the $C_2$ eigenvalues of the site symmetry group of Kramers pairs is always $+i$, $-i$, which exhausts the representations, all the invariants due to $C_2$ are trivial. 

\section{Corner charge classification of the layer groups}
\label{sec:layergroupcornerclassification}

We consider the 80 layer groups labeled in Ref.~\onlinecite{Aroyo2011183} (and available at~\url{http://www.cryst.ehu.es/cgi-bin/subperiodic/programs/nph-sub_gen?subtype=layer&from=table}). First we drop all layer groups that involve nonsymmorphic symmetries, since these are broken by any finite geometry with corners. Then we acknowledge that in some groups, only a subgroup is responsible for quantizing corner charges to fractional values, while the remaining symmetry operations at most pose constraints on the sample geometry and corner charge localization. The corner charge classification of these groups is therefore already determined by a minimal set $\mathrm{S}$ of layer groups that covers all possible ways of enforcing quantization. This set and its classification are given by Table~\ref{tab:relevantlayergroups}.

\begin{table}[H]
\centering
\begin{tabular}{|l|l|l|l|l|}
\hline
group & generators & classification & $Q_\mathrm{c} \mod 2$& same classification \\
\hline
1     &  -          & $\mathbb{Z}_1$ &$\{0\}$            & 4, 5, 8, 9, 10, 11, 12, 13, 27, 28, 29, 30, 31, 32, 33, 34, 35, 36               								\\ \hline
2     & $I$        & $\mathbb{Z}_2$ &$\{0,1\}$  & 7, 14, 15, 16, 17, 18, 39, 43, 44, 45, 46, 52, 62, 64                  									\\ \hline
3     & $C_2^z$    & $\mathbb{Z}_2$  &$\{0,1\}$ & 19, 20, 21, 24, 25                    													 \\ \hline
6     & $I, C_2^z$ & $\mathbb{Z}_2$ &$\{0,1\}$  & 40               																\\ \hline
22     & $C_2^x, C_2^y$ & $\mathbb{Z}_2$&$\{0,1\}$  &                  								   \\ \hline
23     & $M_x, M_y$ & $\mathbb{Z}_2$ &$\{0,1\}$ & 26                 								   \\ \hline
37     & $M_x, M_y, M_z$ & $\mathbb{Z}_2$ &$\{0,1\}$ & 47                					 			  \\ \hline
38     & $M_x, I$ & $\mathbb{Z}_2$ &$\{0,1\}$ & 41, 42, 48                					 	  		\\ \hline
49    & $C_4^z$ & $\mathbb{Z}_4$ &$\{0,1/2,1,3/2\}$             & 52, 54, 56                   															\\ \hline
50    & $C_4^z I$ & $\mathbb{Z}_4$ &$\{0,1/2,1,3/2\}$             & 58, 60                   																 \\ \hline
51    & $C_4^z, I$ & $\mathbb{Z}_4$  &$\{0,1/2,1,3/2\}$            & 63                   																 \\ \hline
53    & $C_4^z, C_2^x, C_2^y$ & $\mathbb{Z}_4$  &$\{0,1/2,1,3/2\}$            & 62                   														 \\ \hline
55    & $C_4^z, M_x, M_y$ & $\mathbb{Z}_4$  &$\{0,1/2,1,3/2\}$            & 64                   														 \\ \hline
57    & $C_4^z I, C_2^x, C_2^y$ & $\mathbb{Z}_4$ &$\{0,1/2,1,3/2\}$             &                    														 \\ \hline
59    & $C_4^z I, M_x, M_y$ & $\mathbb{Z}_4$ &$\{0,1/2,1,3/2\}$             &                    															 \\ \hline
61    & $C_4^z, I, C_2^x, C_2^y$ & $\mathbb{Z}_4$ &$\{0,1/2,1,3/2\}$             &                    														 \\ \hline
65    & $C_3^z$ & $\mathbb{Z}_3$   &$\{0,2/3,4/3\}$          &                   																	 \\ \hline
67    & $C_3^z, C_2^x$ & $\mathbb{Z}_3$  &$\{0,2/3,4/3\}$            & 68                															 \\ \hline
69    & $C_3^z, M_x$ & $\mathbb{Z}_3$  &$\{0,2/3,4/3\}$            & 70                																 \\ \hline
71    & $C_3^z, M_x, I$ & $\mathbb{Z}_6$ &$\{0,1/3,2/3,1,4/3,5/3\}$            & 72                   															 \\ \hline
73    & $C_6^z$           & $\mathbb{Z}_6$  &$\{0,1/3,2/3,1,4/3,5/3\}$           &                    																 \\ \hline
74    &  $C_6^z I$          & $\mathbb{Z}_6$  &$\{0,1/3,2/3,1,4/3,5/3\}$            &                    															 \\ \hline
75    & $C_6^z, I$           & $\mathbb{Z}_6$  &$\{0,1/3,2/3,1,4/3,5/3\}$            &                    														 	\\ \hline
76    & $C_6^z, C_2^x$           & $\mathbb{Z}_6$   &$\{0,1/3,2/3,1,4/3,5/3\}$           &                    														 \\ \hline
77    & $C_6^z, M_x$           & $\mathbb{Z}_6$   &$\{0,1/3,2/3,1,4/3,5/3\}$          &                    															 \\ \hline
78    & $C_6^z I, M_x$           & $\mathbb{Z}_6$  &$\{0,1/3,2/3,1,4/3,5/3\}$            & 79                  														 \\ \hline
80    & $C_6^z, I, M_x$           & $\mathbb{Z}_6$   &$\{0,1/3,2/3,1,4/3,5/3\}$           &                   														 \\ \hline
\end{tabular}
\caption{\label{tab:relevantlayergroups} Corner charge classification and topological indices of $\mathrm{S}$. The boundary classification of any layer group $\mathrm{l}$ is given by that of the group $\mathrm{s} \in \mathrm{S}$, where $\mathrm{s}$ is the largest possible subgroup of $\mathrm{l}$ contained in $\mathrm{S}$. In the case where $\mathrm{l}$ contains nonsymmorphic operations, its classification is the same as that of the layer group $\mathrm{l}'$ that consists of the symmorphic part of $\mathrm{l}$.}
\end{table}

\section{Band representations}
\label{sec:EBRsmat}
We present the band representations of the space groups 164 (P$\bar{3}$m1) and 191 (P6/mmm) relevant for proposed material candidates. To deduce Wyckoff positions from which EBRs can be induced, we use data collected from the Bilbao Crystallographic Server~\cite{Aroyo2011183,Bilbao1, Bradlyn17, Vergniory17}. Note that we discard Wyckoff positions with nonzero $z$-component as they are irrelevant for a 2D geometry.

\begin{table}[H]
	\centering
	\begin{tabular}{|c|c|c|c|}
		\hline 
		SG & phase & band representation & EBRs \\
		\hline
		164 & TI 1 & $(3\bar{\Gamma}_{8}\oplus 2\bar{\Gamma}_{9},2\bar{M}_{3}\bar{M}_{4}\oplus3\bar{M}_{5}\bar{M}_{6},2\bar{K}_{4}\bar{K}_{5}\oplus 3\bar{K}_{6})$ & -  \\
		\hline
		164 & TI 2 & $(2\bar{\Gamma}_{8}\oplus 2\bar{\Gamma}_{9}\oplus\bar{\Gamma}_{4}\bar{\Gamma}_{5},2\bar{M}_{3}\bar{M}_{4}\oplus3\bar{M}_{5}\bar{M}_{6},2\bar{K}_{4}\bar{K}_{5}\oplus 3\bar{K}_{6})$ & -  \\
		\hline
		164 & $3c$ OAL & $(3\bar{\Gamma}_{8}\oplus\bar{\Gamma}_{9}\oplus\bar{\Gamma}_{4}\bar{\Gamma}_{5},2\bar{M}_{3}\bar{M}_{4}\oplus3\bar{M}_{5}\bar{M}_{6},2\bar{K}_{4}\bar{K}_{5}\oplus 3\bar{K}_{6})$ &  $\bar{E}_{1}(2d)  \oplus\ {}^{1}\bar{E}_{g}^{2}\bar{E}_{g}(3c)  $  \\
		\hline
		164 & $1a$ OAL & $(2\bar{\Gamma}_{8} \oplus 2\bar{\Gamma}_{9} \oplus \bar{\Gamma}_{4} \bar{\Gamma}_{5}, 3 \bar{M}_3 \bar{M}_4 \oplus 2 \bar{M}_5 \bar{M_6}, 2 \bar{K}_4 \bar{K}_5 \oplus 3 \bar{K}_6)$ & $\bar{E}_{1g}(1a)  \oplus \bar{E}_{1u}(1a) \oplus \bar{E}_{1}(2d)  \oplus {}^{1}\bar{E}_{g}^{2}\bar{E}_{g}(1a) $ \\
		\hline \hline 
		191 & TCI  & $(\bar{\Gamma}_{7}\oplus \bar{\Gamma}_{8}\oplus\bar{\Gamma}_{9}\oplus\bar{\Gamma}_{11}\oplus \bar{\Gamma}_{12},3\bar{M}_{5}\oplus 2\bar{M}_{6},2\bar{K}_{7}\oplus \bar{K}_{8}\oplus 2\bar{K}_{9})$  & -  \\

		\hline
	\end{tabular} 
	\caption{Band representations corresponding to distinct phases as shown in Fig.~\ref{fig:materials} in the main text. `-' indicates that a given band representation cannot be written as a combination of EBRs.}
	\label{tab:band_reps}
\end{table}

\section{Details of the ab-initio calculations}
\label{sec:DFT}
Fully relativistic DFT calculations were performed via the Vienna \emph{ab initio} simulation package (VASP)~\cite{KresseFurth96-1, KresseFurth96-2} by employing the Perdew-Burke-Ernzerhof (PBE)~\cite{PBE-1, PBE-2} exchange-correlation functional and projected augmented-wave pseudopotentials~\cite{PAW-1, PAW-2}. For the self-consistent calculations, we used a $19 \times 19 \times 1$ $\bs{k}$-point grid generated for the Monkhorst-Pack method in case of Bi and Sb, and a $17 \times 17 \times 1$ mesh for As. The plane wave basis cutoff was set to 400 eV (Bi and Sb) or 350 eV (As). A finer grid of $30 \times 30 \times 1$ $\bs{k}$-points was used later on in order to obtain the energy gaps and band representations. The lattice parameters in the equilibrium configuration, which are in good agreement with previous reports~\cite{LattParamBi, LattParamSb, LattParamAs}, are summarized in Table~\ref{tab:latt_param}.
\begin{table}[H]
	\centering
	\begin{tabular}{|c|c|c|c|}
		\hline 
		& Bi & Sb & As\\
		\hline
		$a$ [$\rm{\AA}$] & 4.39 & 4.04 & 3.61 \\
		\hline 
		$d_{z}$ [$\rm{\AA}$] & 1.74 & 1.65 & 1.40\\ 
		\hline
	\end{tabular} 
	\caption{Lattice constant and buckling parameter for the unstrained (free-standing) buckled structures.}
	\label{tab:latt_param}
\end{table}

For open flake calculations, we employed the Siesta code~\cite{Soler_2002}. We used pseudo-atomic orbitals (PAO) with a basis of double zeta plus polarization orbitals (DZP) and norm-conserving fully relativistic pseudopotentials from the PseudoDojo library~\cite{VANSETTEN201839}. The bulk crystal structure was terminated to obtain a hexagonal structure of 546 Sb atoms, and 30 Te atoms were added to the edges in order to passivate the edge states (as shown in Fig.~\ref{fig:materials}~\textbf{f}). The distance between Te and edge Sb atoms was set to a value 3.02 \AA, which was determined from the structure relaxation of an armchair Sb ribbon with Te adatoms at the edge. The DFT data post-processing was performed with the sisl Python package~\cite{zerothi_sisl}.

The irreducible representations of bands at high-symmetry points were obtained using the \texttt{irrep} code~\cite{irrep-github}, which relies on the double space group character tables~\cite{elcoro2017} published on the Bilbao Crystallographic server~\cite{bilbao-server}.

\end{widetext}

\bibliography{Ref-Lib}

\end{document}